\documentclass[reprint,twocolumn,secnumarabic,amssymb,aps,prx,superscriptaddress,nofootinbib,longbibliography,floatfix]{revtex4-2}

\usepackage{array}[=2016-10-06]
\usepackage{amsmath}
\usepackage{amssymb}
\usepackage{amsfonts}
\usepackage{xcolor}
\usepackage[normalem]{ulem}
\usepackage{graphicx}
\usepackage{bm}
\usepackage{physics}
\usepackage{hyperref}
\hypersetup{
    colorlinks=true,
    linkcolor=black,
    citecolor=black,
    urlcolor=black
}
\usepackage{booktabs}
\usepackage[utf8]{inputenc}
\usepackage{overpic}
\usepackage{multirow}
\usepackage{tabularray}
\usepackage{placeins}
\usepackage{pdfpages}
\makeatletter
\AtBeginDocument{\let\LS@rot\@empty}
\makeatother

\graphicspath{{./}{figures/}}

\makeatletter
\newcommand{\DeclareMethodsRef}[2]{\expandafter\def\csname methodsrefname@#1\endcsname{#2}}
\newcommand{\methodname}[1]{\@ifundefined{methodsrefname@#1}{\textbf{[unknown Methods reference: #1]}}{\hyperref[#1]{\csname methodsrefname@#1\endcsname}}}
\newcommand{\methodsref}[1]{the ``\methodname{#1}'' subsection of Methods}
\makeatother

\DeclareMethodsRef{app:heterogeneous_noise}{Hardware-Inferred Circuit-Level Noise Model}
\DeclareMethodsRef{app:memory_fit}{Surface Code Memory Benchmark}
\DeclareMethodsRef{app:fold_transversal_gates}{Fold-transversal S gates with GRAIL}
\DeclareMethodsRef{app:logical_check}{Logical Check in Magic State Cultivation Compiled into GRAIL}

\makeatletter
\renewcommand{\p@subsection}{}
\renewcommand{\@caption@fignum@sep}{\enspace\textbar\enspace}
\makeatother

\renewcommand{\figurename}{Fig.}
\renewcommand{\tablename}{Table}
\renewcommand{\thetable}{\arabic{table}}

\begin{document}

\title{A Universal Fault-Tolerant Quantum Computing Architecture  with Globally Driven Atom Arrays}

\author{Zhihao Zhang}
\thanks{These authors contributed equally to this work}
\affiliation{State Key Laboratory of Surface Physics, Institute of Nanoelectronics and Quantum Computing, and Department of Physics, Fudan University, Shanghai 200433, China}

\author{Shuaifan Cao}
\thanks{These authors contributed equally to this work}
\affiliation{State Key Laboratory of Surface Physics, Institute of Nanoelectronics and Quantum Computing, and Department of Physics, Fudan University, Shanghai 200433, China}

\author{Xinyang Yu}
\affiliation{State Key Laboratory of Surface Physics, Institute of Nanoelectronics and Quantum Computing, and Department of Physics, Fudan University, Shanghai 200433, China}

\author{Yanyan Chen}
\affiliation{State Key Laboratory of Surface Physics, Institute of Nanoelectronics and Quantum Computing, and Department of Physics, Fudan University, Shanghai 200433, China}

\author{Yueyang Min}
\affiliation{State Key Laboratory of Surface Physics, Institute of Nanoelectronics and Quantum Computing, and Department of Physics, Fudan University, Shanghai 200433, China}

\author{Yingzhou Li}
\affiliation{School of Mathematical Sciences, Fudan University, Shanghai 200433, China}
\affiliation{Shanghai Key Laboratory for Contemporary Applied Mathematics, Fudan University, Shanghai 200433, China}

\author{Zhongcheng Yu}
\affiliation{State Key Laboratory of Surface Physics, Institute of Nanoelectronics and Quantum Computing, and Department of Physics, Fudan University, Shanghai 200433, China}

\author{Lei Feng}
\email{leifeng@fudan.edu.cn}
\affiliation{State Key Laboratory of Surface Physics, Institute of Nanoelectronics and Quantum Computing, and Department of Physics, Fudan University, Shanghai 200433, China}
\affiliation{Shanghai Buchou Quantum Technology Co., Ltd}

\author{Xiaopeng Li}
\email{xiaopeng\underline{ }li@fudan.edu.cn}
\affiliation{State Key Laboratory of Surface Physics, Institute of Nanoelectronics and Quantum Computing, and Department of Physics, Fudan University, Shanghai 200433, China}
\affiliation{Shanghai Buchou Quantum Technology Co., Ltd}

\date{November 9, 2026}

\begin{abstract}

Scalable quantum computation requires reconciling programmability with simple, high-fidelity control. Here, we propose a universal fault-tolerant architecture using globally driven dual-species atom arrays. 
In a hybrid neutral-atom system, interspecies Rydberg interactions enable qubit addressability without locally addressed quantum gates.  We construct a universal native instruction set, Global Rydberg Atom Instruction Language (GRAIL), and show that crosstalk errors can be suppressed to the $10^{-4}$ level. We compile surface-code error correction and magic state cultivation into GRAIL, and parallelize the compilation to suppress crosstalk.  Under a hardware-inferred circuit-level noise model, we obtain an error-correction threshold of \(1.43\% \) for the resulting quantum memory and a logical error rate below \(10^{-9}\) for magic states cultivated at an experimentally accessible Rydberg Rabi frequency. 
By replacing local gate addressing with atom-mediated selectivity, this architecture removes the associated calibration requirements, offering a route to simpler control for large-scale fault-tolerant neutral-atom computation. 

\end{abstract}

\maketitle

Neutral-atom platforms have made rapid progress towards fault-tolerant quantum computing (FTQC)~\cite{Wilk2010RydbergEntanglement,Levine2018RydbergControl,Bluvstein2024,Bluvstein2026,2026qca,Weimer2010RydbergSimulator,Petrosyan2024DualSpeciesGates,Beterov2015DifferentSpecies,2024np,2026saffman,rbcs2022}, with gate fidelities exceeding the thresholds of several quantum error-correcting codes \cite{highfidelity,PRXQuantum.6.010331,PRXQuantum.6.020334,lukin2026highfidelity}.
Experiments have demonstrated both globally driven~\cite{Levine2019Parallel,highfidelity} and locally addressed quantum gates \cite{Isenhower2010CNOT,Maller2015RydbergCNOT,Graham2022NeutralAtomAlgorithms,2025saffman}, while coherent atom transport using optical tweezers~\cite{Beugnon2007Transport,Bluvstein2022CoherentTransport} offers flexible connectivity for implementing high-rate quantum low-density parity-check (qLDPC) codes \cite{Xu2024, Bhardwaj2026HighRateQLDPC}. 
Parallel gate operations enabled by crossed acousto-optic deflectors further support the efficient execution of quantum error correction and quantum algorithms \cite{Levine2019Parallel,highfidelity,Bluvstein2024,Bluvstein2026}. 
Together, these advances have motivated resource estimates and architectural roadmaps~\cite{Zhou2025FullStackDesign, Zhou2025TransversalResources,roadmap2026} for applications ranging from cryptographically relevant computations~\cite{Gidney2021RSA2048,Cain2026ShorAtomicQubits} to large-scale electronic-structure simulations \cite{Reiher2017ReactionMechanisms,Low2025SpectralAmplification}. 
They also sharpen a central question: how can the control capabilities demonstrated in current devices be extended to the scale required for practical FTQC? 

A key challenge is the scalable implementation of programmable, high-fidelity parallel gates~\cite{Graham2023MultiscaleAddressing,Zhang2024ScaledLocalControl,Menssen2023PhotonicControl}.  Architectures based on local optical addressing must deliver precisely controlled fields to target qubits while suppressing unwanted excitation and crosstalk~\cite{address2015,Graham2023MultiscaleAddressing,Zhang2024ScaledLocalControl}. 
Maintaining high fidelity across different addressing patterns requires careful fine-tuning of spatially varying optical fields~\cite{Bluvstein2024,Zhang2024ScaledLocalControl}, with a potentially substantial calibration burden as arrays grow and gate schedules become more complex. 
Global driving alone can also support universal quantum computing~\cite{PhysRevLett.131.170601},  which offers an alternative to the local-addressing scheme. 
Realizing a fault-tolerant architecture, however, requires integrating programmable operations, syndrome extraction and logical gates while controlling the crosstalk induced by global pulses. The central architectural challenge is therefore to combine the control simplicity of global driving with the selectivity and error resilience required for large-scale fault-tolerant quantum computing.

Here, we propose a universal fault-tolerant quantum computing architecture based on a dual-species atom array driven entirely by global control fields, without locally focused beams for gate operations. 
One atomic species encodes the physical qubits, while the other serves as an auxiliary register for controlling operations and probing qubit states. 
Interspecies Rydberg interactions provide qubit addressability through the spatial arrangement of auxiliary atoms, removing the need for locally focused gate beams. 
This approach exploits the identical internal structure and the {\it digital} nature of atoms to reduce reliance on the calibration of locally addressed optical fields.
Specifically, we implement this dual-species architecture using a hybrid rubidium--caesium (Rb--Cs) tweezer array and introduce the Global Rydberg Atom Instruction Language (\(\mathsf{GRAIL}\)). With this language, we construct a universal physical gate set, \(\{H,T,\mathrm{CZ}\}\). 
Using parameters accessible in current experiments, we predict fidelities above \(99.5\%\) for all compiled physical gates, with suppressed crosstalk, and derive a corresponding circuit-level noise model.

Building on this ability to apply gates selectively to physical qubits under global driving, we develop fault-tolerant protocols for storing and processing quantum information. For quantum memory, we compile the ZXXZ surface code~\cite{Fowler2012SurfaceCodes,Google2023} into \(\mathsf{GRAIL}\). 
Under the circuit-level noise model, we obtain a fitted error threshold of $1.43\%$ using minimum-weight perfect matching decoding~\cite{pymatching}. 
Beyond quantum memory, we develop protocols for fault-tolerant logical computation, including transversal implementations of \(\bar H\), \(\bar S\), and \(\overline{\mathrm{CZ}}\)~\cite{logicals2026}, together with non-transversal logical \(\bar T \) gates enabled by magic state cultivation~\cite{2024msc,Chen2026RPCultivation,claes2025msc,p8tw-6kq9,sahay2026foldmsc}. 
These components establish a route to universal FTQC using globally driven dual-species atom arrays. 
More broadly, the architecture separates programmable qubit addressability from local quantum control, offering a potential route to reducing the control complexity of large-scale neutral-atom processors. 
The same principle may extend to high-rate qLDPC codes when supplemented by long-range connectivity through coherent atom transport~\cite{Bluvstein2022CoherentTransport,Xu2024,Bhardwaj2026HighRateQLDPC} or microwave-cavity coupling \cite{Sarkany2015MicrowaveCavity, Chen2026MicrowaveCatBus}.

\section{Universal Physical Gates under Global Control} \label{sec:implementation}

\begin{figure*}[htbp]
    \centering
    \includegraphics[width=\textwidth]{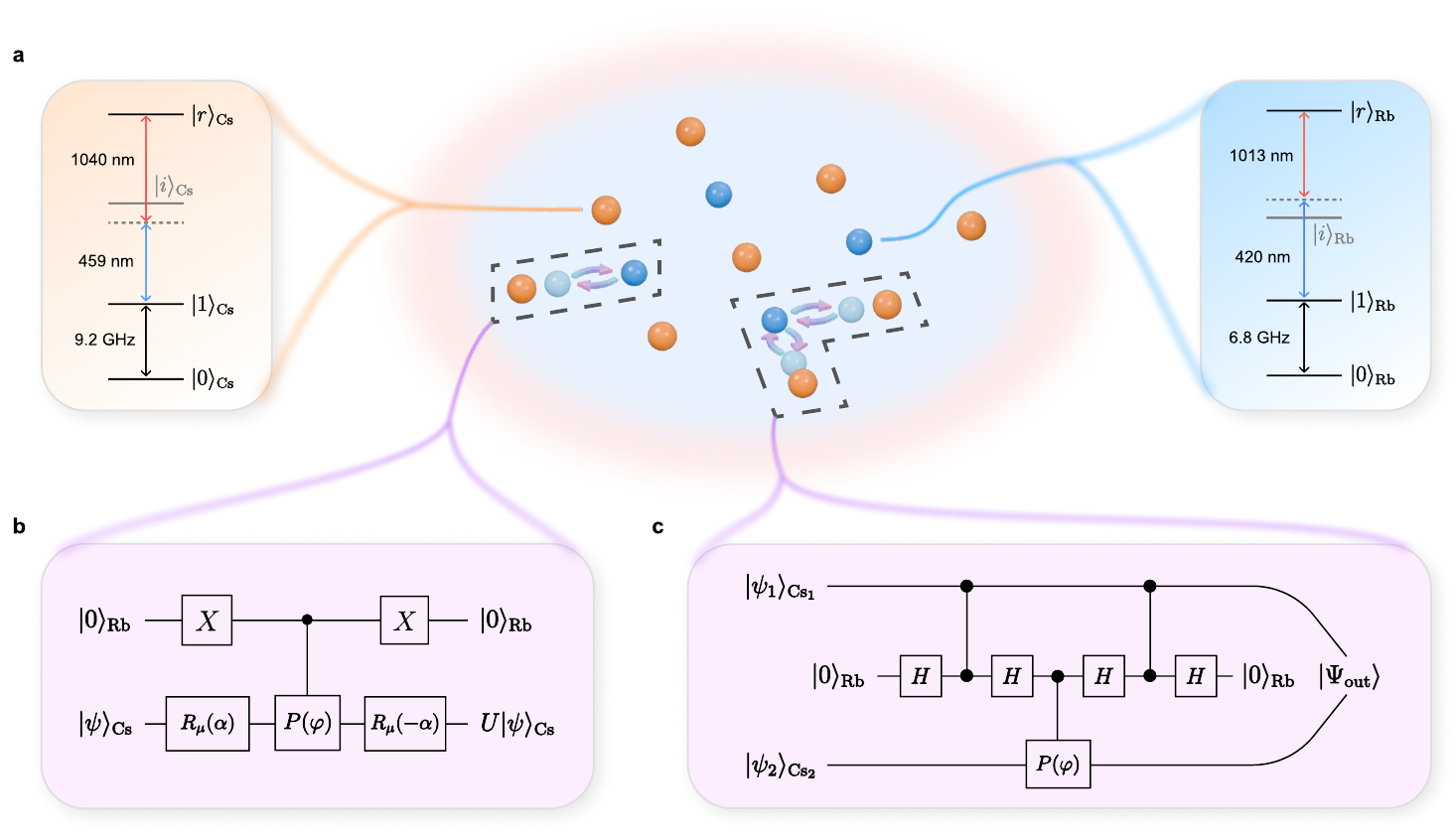}
    \caption{\textbf{Dual-species architecture for globally driven quantum computing.}
    \textbf{a}, Array layout and atomic levels of stationary \(^{133}\mathrm{Cs}\) data qubits (orange) and mobile \(^{87}\mathrm{Rb}\) atoms (blue).
    The Rb atoms serve as addressors for physical gates on Cs qubits or as measure qubits for syndrome extraction.
    Global pulses act on all atoms of the corresponding species, as indicated by blue (Rb) and orange (Cs) halos; arrows show Rb motion.
    The clock states \(\lvert0\rangle,\lvert1\rangle\) are \(\lvert F=3,m_F=0\rangle,\lvert F=4,m_F=0\rangle\) in Cs and \(\lvert F=1,m_F=0\rangle,\lvert F=2,m_F=0\rangle\) in Rb.
    We choose \(\lvert r\rangle_{\mathrm{Cs}}=\lvert67S_{1/2}\rangle\) and \(\lvert r\rangle_{\mathrm{Rb}}=\lvert68S_{1/2}\rangle\) as the Rydberg states for the blockade interaction.
    \textbf{b}, Individually addressed single-qubit gates under global control. 
    An Rb addressor is positioned within the blockade radius of the target Cs qubit to achieve qubit addressability. 
    \textbf{c}, Rb-mediated Cs--Cs controlled-phase gates.
    An Rb addressor sequentially interacts with the two Cs qubits, mediating an effective controlled-phase gate between them while returning to its original state disentangled from the data qubits. }
    \label{fig:architecture}
\end{figure*}

We realize universal quantum computing on a dual-species platform under global control. 
In our architecture, stationary \(^{133}\mathrm{Cs}\) atoms form the data register, while mobile $^{87}\mathrm{Rb}$ atoms serve either as syndrome-extraction (SE) ancillas or as atomic addressors to select Cs qubits. 
For both species, qubits are encoded in ground-state hyperfine clock states, while native controlled-phase gates use the van der Waals interaction between the Cs \(\lvert67S_{1/2}\rangle\) and Rb \(\lvert68S_{1/2}\rangle\) Rydberg states shown in Fig.~\ref{fig:architecture}\textbf{a}.
Moving an Rb addressor within the blockade radius of a selected Cs qubit provides spatial addressing under global microwave and Rydberg pulse sequences. Reconfiguring the Rb array in this way eliminates the need to implement and calibrate focused Raman beams at individual Cs sites~\cite{Bluvstein2024,Bluvstein2026}. Building on this physical architecture, we introduce the native instructions as follows.

Global microwave pulses drive high-fidelity single-qubit rotations in neutral-atom arrays~\cite{address2015,Sheng2018Microwave,Nikolov2023Microwave,Manetsch2025}. 
In our dual-species Rb--Cs system, the well-separated hyperfine transition frequencies enable independent global microwave control of the two species. 
We use \(s\in\{\mathrm{Rb},\mathrm{Cs},\mathrm{RbCs}\}\) to specify whether a global operation acts on Rb atoms, Cs atoms, or both species.
Let \(\mathcal A_{\mathrm{Rb}}\) and \(\mathcal A_{\mathrm{Cs}}\) denote the complete sets of atoms of the two species, with \(\mathcal A_{\mathrm{RbCs}}=\mathcal A_{\mathrm{Rb}}\cup\mathcal A_{\mathrm{Cs}}\) and \(N_s=|\mathcal A_s|\). 
A global microwave rotation applies the same single-qubit unitary \(U\) to all atoms of a chosen species,
\begin{equation}
G_s(U)=\bigotimes_{i\in\mathcal A_s}U_i,
\label{eq:native_global_rotation}
\end{equation}
where $U_i$ denotes $U$ on atom $i$.

Coherent atom transport with optical tweezers has been demonstrated in neutral-atom arrays~\cite{Beugnon2007Transport,Bluvstein2022CoherentTransport,Manetsch2025,Bluvstein2026}. 
In our architecture, Rb atoms are transported over finite distances using optical tweezers controlled by acousto-optic deflectors (AODs), while Cs atoms remain stationary. Rb shuttling provides addressability under global driving by selecting which atom pairs lie within the Rydberg blockade radius. We specify the intended Rb--Rb and Rb--Cs pairs by
\begin{equation}
\mathcal Q\subseteq
\{(i,j)\mid
i\in\mathcal A_{\mathrm{Rb}},
\ j\in\mathcal A_{\mathrm{Rb}}\cup\mathcal A_{\mathrm{Cs}},
\ i\ne j\}.
\end{equation}
The pairs in \(\mathcal Q\) are disjoint: each atom appears in at most one pair.
The shuttling instruction \(\mathrm{Sh}_{\mathrm{Rb}}(\mathcal Q)\) moves each Rb atom \(i\) from its current position to the neighbourhood of its partner \(j\) (within the blockade radius).
We define the inverse operation, \(\mathrm{Sh}^{-1}_{\mathrm{Rb}}(\mathcal Q)\), as the reverse movement of the Rb atoms back to their original positions.

Global Rydberg pulses can generate entanglement for those atoms within the Rydberg blockade radius~\cite{Levine2019Parallel,highfidelity,2024np}.
In our architecture, we apply a global Rydberg pulse \(\mathrm{Ryd}_s(\varphi)\) to the atomic species designated by $s$. 
The designed global Rydberg pulse implements a native controlled-phase gate $\mathrm{CP}(\varphi)$ on all atom pairs within the Rydberg blockade radius (Supplementary Note~1).

Species-selective state preparation and measurement can be performed in dual-species neutral-atom arrays~\cite{rbcs2022,2024np}. 
The widely separated optical transitions of Rb and Cs allow high-fidelity reset and readout of one species with negligible crosstalk~\cite{2023science}.
In our architecture, global reset and measurement act on all atoms in \(\mathcal A_s\) and are performed directly without extra atom movement.
The reset \(\mathcal R_s\) prepares these atoms in \(\lvert0\rangle\), giving the state
\(
\lvert0\rangle^{\otimes N_s}
\).
The measurement \(\mathcal M_s\) reads out these atoms in the computational basis and returns the outcome.

Taken together, the five instruction families above form a native instruction set architecture, dubbed Global Rydberg Atom Instruction Language ($\mathsf{GRAIL}$),
\begin{equation}
\mathsf{GRAIL}
=
\{
G_s(U),\,
\mathrm{Sh}^{\pm1}_{\mathrm{Rb}}(\mathcal Q),\,
\mathrm{Ryd}_s(\varphi),\,
\mathcal R_s,\,
\mathcal M_s
\}.
\label{eq:native_instruction_set}
\end{equation}

In $\mathsf{GRAIL}$, atom shuttling and global Rydberg pulses can be combined to implement a native controlled-phase gate on selected atom pairs defined by \(\mathcal Q\), \begin{equation}
    \begin{aligned}
    \mathrm{CP}(\mathcal Q;\varphi) & \equiv \prod_{(i,j)\in\mathcal Q}\mathrm{CP}_{i,j}(\varphi) \\
    & =\mathrm{Sh}^{-1}_{\mathrm{Rb}}(\mathcal Q)\mathrm{Ryd}_{s}(\varphi)\mathrm{Sh}_{\mathrm{Rb}}(\mathcal Q),
    \end{aligned}
\label{eq:native_cp_layer}
\end{equation}
where
\(\mathrm{CP}_{i,j}(\varphi)
=\operatorname{diag}(1,1,1,e^{i\varphi})\)
in the computational basis, and the involved atomic species specified by $s$ is determined by \(\mathcal Q\).
Although the Rydberg pulse is global, only Cs atoms paired with Rb atoms acquire the conditional phase.

Building on $\mathsf{GRAIL}$, we first realize single-qubit physical gates on chosen Cs qubits.
Any single-qubit unitary $U$ can be decomposed using a rotation $R_\mu (\alpha)$ around the $\mu$ axis with angle $\alpha$ and a phase gate \(P(\varphi)\) as
\begin{equation}
\begin{aligned}
&U=R_\mu (-\alpha) P(\varphi)R_\mu (\alpha),\\
&P(\varphi)=\operatorname{diag}(1,e^{i\varphi}).
\end{aligned}
\label{eq:selected_cs_unitary}
\end{equation}
Let \(\mathcal T_{\mathrm{Cs}}\subseteq\mathcal A_{\mathrm{Cs}}\) denote the set of target Cs qubits.
For each target \(j\in\mathcal T_{\mathrm{Cs}}\), we allocate a distinct neighbouring spare Rb atom \(a(j)\in\mathcal A_{\mathrm{Rb}}\), prepared in \(\lvert0\rangle\), as its addressor.
The resulting pair set is \(\mathcal Q=\{(a(j),j)\mid j\in\mathcal T_{\mathrm{Cs}}\}\).
The $\mathsf{GRAIL}$ instruction sequence implementing $U$ on \(\mathcal T_{\mathrm{Cs}}\), as illustrated in Fig.~\ref{fig:architecture}\textbf{b}, is 
\begin{equation*}
G_{\mathrm{Rb}}(X)G_{\mathrm{Cs}}\big(R_\mu (-\alpha)\big)\mathrm{CP}(\mathcal Q;\varphi)G_{\mathrm{Cs}}\big(R_\mu (\alpha)\big)G_{\mathrm{Rb}}(X).
\end{equation*}
For the physical $H$ and $T$ gates, we have \((\mu,\alpha,\varphi)=(y,-\pi/4,\pi)\) and \((y,0,\pi/4)\), respectively.
In this design, the sequence acts as the identity on idling qubits. 

To realize a physical controlled-phase gate on a specific Cs--Cs pair, we use sequential Rb--Cs interactions for entanglement transfer via a shared Rb addressor.
Let \(\mathcal T^1_{\mathrm{Cs}},\mathcal T^2_{\mathrm{Cs}}\subseteq\mathcal A_{\mathrm{Cs}}\) denote the sets of first and second qubits in the disjoint target Cs--Cs pairs, respectively. 
For each target pair \((j_1,j_2)\), we allocate a neighbouring spare Rb atom \(a(j_1)=a(j_2)\), prepared in \(\lvert0\rangle\), as their shared addressor.
The two shuttling pair sets are \(\mathcal Q^b=\{(a(j),j)\mid j\in\mathcal T^b_{\mathrm{Cs}}\}\), \(b=1,2\).
The $\mathsf{GRAIL}$ instruction sequence implementing the physical controlled-phase gate on $\mathcal{T}_\mathrm{Cs}^1$ and $\mathcal{T}_\mathrm{Cs}^2$, as illustrated in Fig.~\ref{fig:architecture}\textbf{c}, is
\begin{equation*}
    \begin{aligned}
    &G_{\mathrm{Rb}}(H)\mathrm{CP}(\mathcal Q^1;\pi)G_{\mathrm{Rb}}(H)\mathrm{CP}(\mathcal Q^2;\varphi) \\
    & \circ G_{\mathrm{Rb}}(H)\mathrm{CP}(\mathcal Q^1;\pi)G_{\mathrm{Rb}}(H).
    \end{aligned}
\end{equation*}
For the physical CZ gates, we have \(\varphi=\pi\).
In this entangling gate design, this sequence implements a controlled-phase gate on each target Cs--Cs pair and sets its addressor back to \(\lvert0\rangle\), disentangled from the data qubits.
It acts as the identity on idling qubits.
The ideal action of these sequences on the target qubits,
addressors and idling atoms is derived in Supplementary Note~1.

With these constructions, our architecture can implement the universal physical gate set $\{H, T, \mathrm{CZ}\}$ with full qubit addressability by combining global quantum control with coherent atom shuttling.
Sequences of \(\mathsf{GRAIL}\) instructions can therefore approximate arbitrary unitary operations on the Cs data register to any desired precision.

\section{Physical Gate Performance under Global Control}
\label{sec:error_analysis}

To characterize the performance of $\mathsf{GRAIL}$ in actual physical implementation, it is crucial to analyse the gate fidelity on data qubits and the crosstalk on idling qubits.
We focus on the intrinsic limitations imposed by Rydberg-state decay and finite blockade strength, which cause both leakage and phase errors~\cite{Jandura2022TimeOptimal,Pagano2022ErrorBudgeting}.
These errors are therefore primarily governed by the two-photon Rydberg Rabi frequency, $\Omega/2\pi$, which typically ranges from $1$ to $20$ MHz in experiments~\cite{2024np,lukin2026highfidelity}.
We have derived a hardware-inferred circuit-level noise model for our gate realization by incorporating both unitary and dissipative dynamics (Methods, Extended Data Table~\ref{tab:noise_model}).
Here, we estimate gate infidelity and crosstalk for each compiled gate sequence and present the main results below.

Particularly for atom pairs participating in quantum gates, the leakage error is quantified by the leakage probability \(p\equiv p_{\mathrm{leak},\mathrm{RbCs}}^\pi\), defined as the input-averaged probability that at least one atom leaves the computational subspace (Methods).
Taking into account the phase errors in addition, we analyse the performance of the addressed physical \(H\), \(T\) and Cs--Cs CZ gates. 
With \(\Omega/2\pi=2.38\,\mathrm{MHz}\), the infidelities of these gates are \(2.05\times10^{-3}\), \(9.16\times10^{-4}\) and \(4.74\times10^{-3}\), respectively, which are below the threshold of common quantum error correction codes~\cite{Stephens2014SurfaceCodeThresholds,Xu2024,Bhardwaj2026HighRateQLDPC}.

\begin{figure}[t]
    \centering
    \includegraphics[width=\columnwidth]{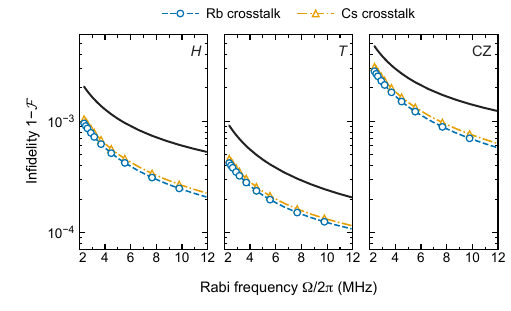}
    \caption{\textbf{Universal physical gate performance under global control.}
    Infidelities of the compiled physical \(H\), \(T\) and \(\mathrm{CZ}\) gates on addressed Cs qubits (black solid curves) are plotted against Rydberg Rabi frequency \(\Omega/2\pi\).
    Crosstalk errors on idling Rb and Cs atoms are quantified by the infidelity $1-\mathcal F_{\mathrm{cr}}$ defined in equation~\eqref{eq:crosstalk_haar_definition}.
    The simulations include Rydberg decay and finite blockade strength \(V_{\mathrm{RbCs}}/\Omega=10\).}
    \label{fig:universal_gate_performance}
\end{figure}

Besides, global quantum control pulses transiently excite idling atoms to Rydberg states, causing unwanted crosstalk.
The crosstalk errors are quantified by the probability of these atoms failing to return to their original states, $1-\mathcal F_{\mathrm{cr}}$, with
\begin{equation}
\mathcal F_{\mathrm{cr}}
=
\mathbb E_{\psi}\!\left[
\langle\psi|
\mathcal E_{\mathrm{cr}}\big(|\psi\rangle\langle\psi|\big)
|\psi\rangle
\right],
\label{eq:crosstalk_haar_definition}
\end{equation}
where \(\mathbb E_{\psi}\) denotes a uniform average over pure single-qubit input states, and \(\mathcal E_{\mathrm{cr}}\) is the quantum channel for the corresponding gate sequence (Methods and Supplementary Note~1).
With \(\Omega/2\pi=12\,\mathrm{MHz}\), the crosstalk errors for the \(H\), \(T\) and CZ are below 0.1\% (Fig.~\ref{fig:universal_gate_performance}).
These errors scale inversely with the Rydberg Rabi frequency and can thus be further suppressed by increasing the driving strength~\cite{2026saffman,lukin2026highfidelity}.

\section{Quantum Memory with GRAIL}
\label{sec:surface_code_memory}

Fault-tolerant quantum memory provides continuous protection of logical quantum information required during universal quantum computing.
Since our quantum computing architecture is universal, all quantum error correction codes can be compiled into \(\mathsf{GRAIL}\). 
As an example, we demonstrate the compilation for the ZXXZ surface code~\cite{xzzx2021,Google2025} and evaluate its fault-tolerant performance under our hardware-inferred circuit-level noise model.

\begin{figure*}[t]
    \centering
    \includegraphics[width=\textwidth]{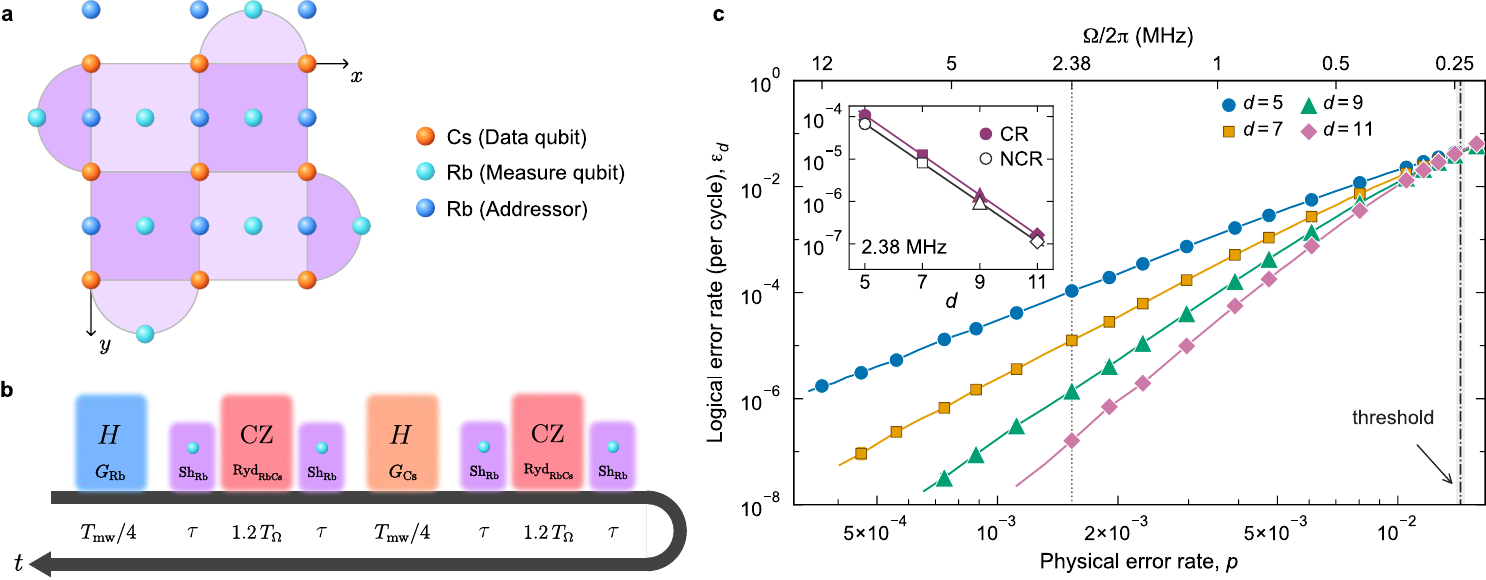}\caption{\textbf{ZXXZ surface code memory under global control.}
    \textbf{a}, A code layout (\(d=3\)) with Cs data qubits, Rb measure qubits, \(X\)-type stabilizers (dark purple), \(Z\)-type stabilizers (light purple), and Rb addressors.
    Rb addressors mediate the individually addressed \(H\) operations for logical initialization and readout. 
    \textbf{b}, The unitary part of an SE round compiled into $\mathsf{GRAIL}$.
    The second half of this unitary part is the time reverse of the displayed instruction sequence.
    \textbf{c}, Memory performance under the hardware-inferred circuit-level noise model. 
    The logical error rates per cycle \(\varepsilon_d\) are averaged over the \(\bar X\) and \(\bar Z\) memory tests after \(d\) SE rounds.
    The physical error rate \(p\) is the leakage probability of the native Rb--Cs CZ gate. 
    Error bars show 95\% confidence intervals. 
    The threshold is \(p_{\mathrm{th}}=1.43\times10^{-2}\).
    The inset shows how the crosstalk affects the logical error rate at Rydberg Rabi frequency \(\Omega/2\pi=2.38\,\mathrm{MHz}\). 
    CR corresponds to the full hardware-inferred circuit-level noise model while NCR shows the results of removing crosstalk errors. 
    The error suppression factors are \(\Lambda_{\mathrm{CR}}\approx 8.76\) and \(\Lambda_{\mathrm{NCR}}\approx 8.47\).
    The number of independent Monte Carlo shots per data point ranges from \(10^4\) to \(10^9\) across the two logical bases combined.}
    \label{fig:memory_performance}
\end{figure*}

In our architecture, each ZXXZ syndrome-extraction (SE) round, \(\mathcal C_{\mathrm{SE}}\), consists of global Rb reset \(\mathcal R_{\mathrm{Rb}}\), the unitary part (Fig.~\ref{fig:memory_performance}\textbf{b}), and global Rb readout \(\mathcal M_{\mathrm{Rb}}\).
Parallel execution under global control enables each SE round to be implemented with a circuit depth independent of the code distance \(d\) (Methods and Supplementary Note~2).

This simplicity, however, comes with a challenge, as global Rydberg pulses also drive idling atoms.
The number of idling atoms subject to crosstalk increases with \(d\).
To determine whether this compromises the below-threshold behavior, we evaluate the compiled quantum memory using the hardware-inferred circuit-level noise model incorporating both gate errors and crosstalk on Rb and Cs atoms (Methods).
Despite crosstalk, our simulations show that the quantum memory remains fault-tolerant.
It exhibits a threshold of \(p_{\mathrm{th}}=1.43\%\), corresponding to \(\Omega/2\pi=0.242\,\mathrm{MHz}\) (Fig.~\ref{fig:memory_performance}\textbf{c}). Below threshold, the logical error rate per cycle \(\varepsilon_d\) decreases exponentially with code distance.
At \(\Omega/2\pi=2.38\,\mathrm{MHz}\), increasing the code distance by two suppresses \(\varepsilon_d\) by a factor of \(8.76\pm0.04\) (Fig.~\ref{fig:memory_performance}\textbf{c}, inset). 
Motivated by the low logical error rates needed for quantum algorithms~\cite{Viszlai2026QSIEVE,Cain2026ShorAtomicQubits,Gidney2017Factoring,Haner2020EllipticCurve,Babbush2018ElectronicSpectra,Gidney2025RSA2048,Zhou2025TransversalResources}, we extrapolate the fitted trend and estimate that a $d=17$ memory could reach a logical error rate below \(10^{-9}\).
The required atom count is within the range demonstrated in neutral-atom tweezer arrays~\cite{Manetsch2025} (Methods).
These results show that \(\mathsf{GRAIL}\) can support fault-tolerant quantum memory with parameters accessible in current experiments.

\section{GRAIL Compilation of Logical Operations}
\label{sec:architecture_native_logical}

A universal fault-tolerant quantum computer requires the ability to perform logical gates on encoded quantum information. 
We extend our architecture beyond quantum memory by implementing logical Clifford gates and preparing non-Clifford resources via $\mathsf{GRAIL}$. 

\begin{figure*}[!t]
    \centering
    \includegraphics[width=\textwidth]
        {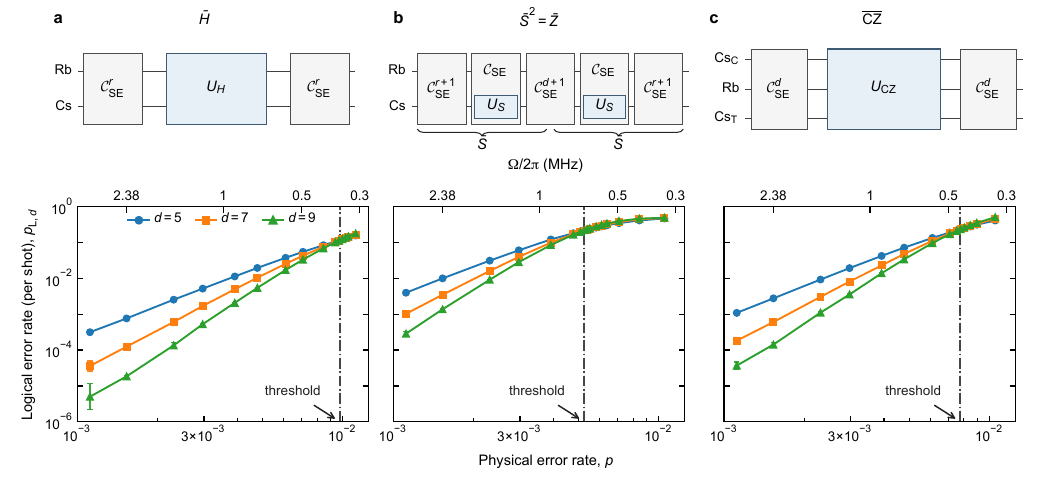}
    \caption{\textbf{Performance of GRAIL-compiled logical Clifford gates.}
    \textbf{a}, \textbf{b} and \textbf{c} show the logical error rates of the \(\bar H\), \(\bar S^2=\bar Z\) and \(\overline{\mathrm{CZ}}\) gates, with thresholds of \(0.97\%\), \(0.52\%\) and \(0.78\%\), respectively.
    Superscripts on \(\mathcal C_{\mathrm{SE}}\) indicate the number of consecutive SE rounds, with \(r=(d+1)/2\).
    Between SE rounds, \(U_H\) implements \(\bar H\) by applying addressed physical \(H\) gates to Cs data qubits,  \(U_S\) inserted halfway through an SE round implements fold-transversal \(\bar S\) and \(U_{\mathrm{CZ}}\) implements \(\overline{\mathrm{CZ}}\) through Rb-mediated CZ gates between corresponding Cs data qubits.
    Error bars show 95\% confidence intervals. 
    The number of independent Monte Carlo shots per data point ranges from \(10^4\) to \(10^9\).}
    \label{fig:logical_clifford_thresholds}
\end{figure*}

For the logical Clifford gates, we first compile transversal \(\bar H\), fold-transversal \(\bar S\) and pairwise transversal \(\overline{\mathrm{CZ}}\) constructions into $\mathsf{GRAIL}$ and integrate them with ZXXZ SE rounds~\cite{logicals2026,Breuckmann2024foldtransversal}.
The logical \(\bar H\) is implemented through an addressed physical \(H\) layer, while the logical \(\overline{\mathrm{CZ}}\) uses parallel Rb-mediated interactions between corresponding data qubits of two code patches.
The fold-transversal \(\bar S\) construction requires additional care: quantum information temporarily held by Rb syndrome ancillas is transferred to Cs storage atoms to accommodate global Rb resets (Methods and Supplementary Note~3).
Under the hardware-inferred circuit-level noise model, the compiled logical \(\bar H\), \(\bar S\) and \(\overline{\mathrm{CZ}}\) exhibit thresholds of \(0.97\%\), \(0.52\%\) and \(0.78\%\), respectively (Fig.~\ref{fig:logical_clifford_thresholds}). At \(\Omega/2\pi=11.9\,\mathrm{MHz}\), extrapolation gives \(d=17\) at which all three logical Clifford gates achieve logical error rates below \(10^{-9}\) (Supplementary Note~3).

\begin{figure*}[!t]
    \centering
    \includegraphics[width=\textwidth]{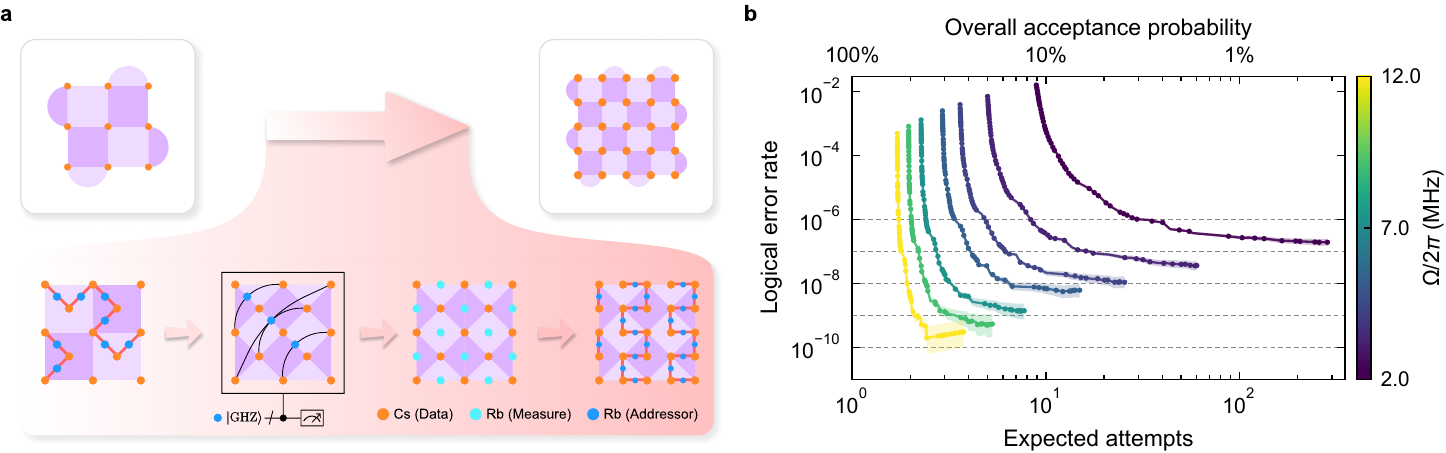}
    \caption{\textbf{Parallelized compilation of magic state cultivation with globally driven atoms.}
    \textbf{a}, Compilation scheme of one cultivation step, illustrated for \(d=3\).
    The rotated surface code \(\mathrm{Rot}(d)\) is converted to the unrotated surface code \(\mathrm{Reg}(d)\), followed by the logical \(\bar H_{XY}\) measurements using Rb GHZ ancillas, completing the growth to \(\mathrm{Rot}(2d-1)\).
    \textbf{b}, Performance of the MSC protocol compiled into \(\mathsf{GRAIL}\) using our parallelized scheme. 
    The logical error rate \(\varepsilon_L\) is plotted against the expected attempts.
    Shaded bands show the statistical uncertainty at the 95\% confidence level. 
    Sampling comprises of order \(10^{10}\) shots for each Rabi frequency.
    }
    \label{fig:architecture_native_msc}
\end{figure*}

Next, we proceed to produce non-Clifford resources via recently proposed magic state cultivation (MSC)~\cite{2024msc,Chen2026RPCultivation,claes2025msc,p8tw-6kq9,sahay2026foldmsc}. Conventional implementations of MSC rely on sequences of locally addressed quantum operations and instantaneous feedback.
Direct compilation into $\mathsf{GRAIL}$ then involves a large fraction of idling atoms and the consequent crosstalk errors compromise the success probability of magic-state preparation. 
To overcome this challenge, we develop a parallelized compilation that reduces the number of global pulses and the crosstalk errors.
As a concrete example, we apply this strategy to the \(f=5\) fold-transversal MSC protocol~\cite{sahay2026foldmsc}, which prepares the logical magic state \(\lvert\bar H_{XY}\rangle=(\lvert\bar0\rangle+e^{i\pi/4}\lvert\bar1\rangle)/\sqrt2\) through two cultivation steps involving fault-tolerant logical \(\bar H_{XY}\) measurements at code distances \(d=3\) and \(d=5\).

To perform such fault-tolerant logical measurements, we need to prepare a GHZ ancilla and construct a controlled-\(\bar H_{XY}\) gate (Fig.~\ref{fig:architecture_native_msc}\textbf{a}).
A parallelized scheme is developed for GHZ ancilla preparation to reduce the global-pulse depth.
A low-depth construction based on local operations and classical communication (LOCC) provides a starting point~\cite{Piroli2021LOCC}.
However, a direct implementation of the original protocol would require mid-circuit measurements of additional preparation ancillas, whereas global Rb readout \(\mathcal M_{\mathrm{Rb}}\) also measures the Rb atoms forming the GHZ ancilla. We overcome this constraint by postponing the preparation-ancilla measurements until after the GHZ-controlled interactions with the Cs data qubits. 
Both groups of Rb ancillas are then measured together. Combined with parallel atomic shuttling, this construction requires only two \(\mathrm{Ryd}_{\mathrm{Rb}}(\pi)\) layers for GHZ preparation at both \(d=3\) and \(d=5\).

After the GHZ ancilla preparation, we next construct the controlled-\(\bar H_{XY}\) operation. 
The controlled-\(\bar X\) component follows from the Rb-mediated gate constructions introduced above, whereas controlled-\(\bar S\) involves multiple controlled-\(S\), controlled-\(S^\dagger\) and CCZ gates~\cite{sahay2026foldmsc}. 
The controlled-\(S\) and controlled-\(S^\dagger\) gates can be implemented directly using the controlled-phase gates introduced in equation~\eqref{eq:native_cp_layer}. 
The main challenge lies in the CCZ gates, for which we develop a decomposition requiring only nine global Rydberg pulse layers (Supplementary Note~1).
Parallel execution allows all required CCZ gates to share these pulse layers, maintaining the same global-pulse depth for both cultivation steps at \(d=3\) and \(d=5\).
Together with the low-depth GHZ preparation, this construction reduces the global-pulse overhead and associated crosstalk during cultivation.

Finally, with our $\mathsf{GRAIL}$ implementation, we evaluate the MSC performance by employing a logical-\(\bar Y\) cultivation benchmark to estimate the performance of the target \(\lvert\bar H_{XY}\rangle\) protocol~\cite{2024msc,sahay2026foldmsc}.
The sampling and postselection procedures are specified in Supplementary Note~4.
The logical error rate \(\varepsilon_L\) varies with the expected attempts, as shown in Fig.~\ref{fig:architecture_native_msc}\textbf{b}.
At a Rydberg Rabi frequency \(\Omega/2\pi=2.38\,\mathrm{MHz}\), we can achieve \(\varepsilon_L=1.94\times10^{-7}\) with \(258\) expected attempts. 
This can be further improved by increasing \(\Omega\): at \(\Omega/2\pi=11.9\,\mathrm{MHz}\), the logical error rate reaches \(1.98\times10^{-10}\) with \(2.43\) expected attempts.
These results substantiate that our parallelized compilation using $\mathsf{GRAIL}$ can efficiently cultivate magic states with only global quantum control.
By supporting both logical Clifford gates and non-Clifford state preparation, \(\mathsf{GRAIL}\) provides a route to universal fault-tolerant quantum computation.

\section{Discussion}
\label{sec:discussion_outlook}

We have developed a globally driven dual-species neutral-atom architecture and designed Rydberg pulse sequences that implement a universal gate set. 
The architecture shifts qubit addressing from calibrated local light fields to controlled Rb positions, while preserving global coherent control across the processor. 
It fully exploits the {\it identical}  internal structure of atoms, and avoids the need to calibrate the {\it analogue} optical fields. 
The resulting native instruction set, $\mathsf{GRAIL}$, contains atom shuttling and global quantum control only. 
This approach offers a route to simplifying the control requirements of large-scale quantum computing, by transforming the demands of delicate qubit-addressing quantum control towards engineering classical control of atom positions. 

Based on this architecture, we establish a concrete route to fault-tolerant computation using the surface code. 
Under the hardware-inferred circuit-level noise model, the quantum memory exhibits an error-correction threshold of $1.43\%$, including the effects of global-pulse-induced crosstalk. 
Logical Clifford operations admit efficient implementations in $\mathsf{GRAIL}$, with fault-tolerant operation predicted for physical parameters compatible with current Rb--Cs technology. 
To enable logical $T$ gates, we adapt magic-state cultivation to the architecture and develop parallelized compilation schemes that suppress idling errors. 
For the experimental parameters considered, our analysis predicts a cultivated-state logical error rate at the $10^{-9}$ level. 

The universality of $\mathsf{GRAIL}$ also motivates extensions beyond the surface code, including high-rate quantum low-density parity-check (qLDPC) codes~\cite{Xu2024,Bhardwaj2026HighRateQLDPC,Zhao2026UltraHighRate,Cain2026ShorAtomicQubits,Bergonzoni2026StaticBuses}.
Realizing these codes efficiently would require implementing their nonlocal connectivity while controlling the associated noise and resource overheads.
Potential approaches include cat-bus protocols~\cite{Chen2026MicrowaveCatBus}, longer-range transport of Rb atoms~\cite{Bluvstein2022CoherentTransport,Zhou2026AlgebraicSurgery}, and teleportation-based operations~\cite{Berthusen2025LocalQLDPC,Shaw2026Networked,Pecorari2025Addressable,Stack2026Transversal,Khan2026EarlyFT}.
Establishing the fault-tolerant performance and space--time costs of these approaches under global control is an important direction for future work.

\clearpage
\bibliographystyle{sn-nature}
\bibliography{references_ncs}

@misc{roadmap2026,
  author = {Adrian J. Menssen and
            Tout Wang and
            Michael Gullans and
            Tom Manovitz and
            Jacob M. Taylor and
            Jason Cong and
            Josiah Sinclair and
            Ziv Aqua and
            Daniel J. Blumenthal and
            Bonilla Ataides, J. Pablo and
            others},
  title = {Strategic Plan for Neutral Atom Quantum Computation},
  year = {2026},
  eprint = {2607.21554},
  archiveprefix = {arXiv},
  primaryclass = {quant-ph},
  url = {https://arxiv.org/abs/2607.21554}
}

@article{highfidelity,
  author = {Simon J. Evered and
            Dolev Bluvstein and
            Marcin Kalinowski and
            Sepehr Ebadi and
            Tom Manovitz and
            Hengyun Zhou and
            Sophie H. Li and
            Alexandra A. Geim and
            Tout T. Wang and
            Nishad Maskara and
            others},
  title = {High-fidelity parallel entangling gates on a neutral-atom quantum computer},
  journal = {Nature},
  volume = {622},
  number = {7982},
  pages = {268--272},
  year = {2023},
  doi = {10.1038/s41586-023-06481-y}
}

@article{Bluvstein2022CoherentTransport,
  author = {Bluvstein, Dolev and
            Levine, Harry and
            Semeghini, Giulia and
            Wang, Tout T. and
            Ebadi, Sepehr and
            Kalinowski, Marcin and
            Keesling, Alexander and
            Maskara, Nishad and
            Pichler, Hannes and
            Greiner, Markus and
            others},
  title = {A quantum processor based on coherent transport of entangled atom arrays},
  journal = {Nature},
  volume = {604},
  number = {7906},
  pages = {451--456},
  year = {2022},
  doi = {10.1038/s41586-022-04592-6}
}

@article{Manetsch2025,
  author = {Manetsch, Hannah J. and
            Nomura, Gyohei and
            Bataille, Elie and
            Lv, Xudong and
            Leung, Kon H. and
            Endres, Manuel},
  title = {A tweezer array with 6,100 highly coherent atomic qubits},
  journal = {Nature},
  volume = {647},
  number = {8088},
  pages = {60--67},
  year = {2025},
  doi = {10.1038/s41586-025-09641-4}
}

@article{Isenhower2010CNOT,
  author = {Isenhower, L. and
            Urban, E. and
            Zhang, X. L. and
            Gill, A. T. and
            Henage, T. and
            Johnson, T. A. and
            Walker, T. G. and
            Saffman, M.},
  title = {Demonstration of a Neutral Atom Controlled-{NOT} Quantum Gate},
  journal = {Phys. Rev. Lett.},
  volume = {104},
  number = {1},
  pages = {010503},
  year = {2010},
  doi = {10.1103/PhysRevLett.104.010503}
}

@article{Wilk2010RydbergEntanglement,
  author = {Wilk, T. and
            Ga{\"e}tan, A. and
            Evellin, C. and
            Wolters, J. and
            Miroshnychenko, Y. and
            Grangier, P. and
            Browaeys, A.},
  title = {Entanglement of Two Individual Neutral Atoms Using {Rydberg} Blockade},
  journal = {Phys. Rev. Lett.},
  volume = {104},
  number = {1},
  pages = {010502},
  year = {2010},
  doi = {10.1103/PhysRevLett.104.010502}
}

@article{Maller2015RydbergCNOT,
  author = {Maller, K. M. and
            Lichtman, M. T. and
            Xia, T. and
            Sun, Y. and
            Piotrowicz, M. J. and
            Carr, A. W. and
            Isenhower, L. and
            Saffman, M.},
  title = {{Rydberg}-blockade controlled-{NOT} gate and entanglement in a two-dimensional array of neutral-atom qubits},
  journal = {Phys. Rev. A},
  volume = {92},
  number = {2},
  pages = {022336},
  year = {2015},
  doi = {10.1103/PhysRevA.92.022336}
}

@article{Levine2018RydbergControl,
  author = {Levine, Harry and
            Keesling, Alexander and
            Omran, Ahmed and
            Bernien, Hannes and
            Schwartz, Sylvain and
            Zibrov, Alexander S. and
            Endres, Manuel and
            Greiner, Markus and
            Vuleti{\'c}, Vladan and
            Lukin, Mikhail D.},
  title = {High-Fidelity Control and Entanglement of {Rydberg}-Atom Qubits},
  journal = {Phys. Rev. Lett.},
  volume = {121},
  number = {12},
  pages = {123603},
  year = {2018},
  doi = {10.1103/PhysRevLett.121.123603}
}

@article{Levine2019Parallel,
  author = {Levine, Harry and
            Keesling, Alexander and
            Semeghini, Giulia and
            Omran, Ahmed and
            Wang, Tout T. and
            Ebadi, Sepehr and
            Bernien, Hannes and
            Greiner, Markus and
            Vuleti\ifmmode \acute{c}\else \'{c}\fi{}, Vladan and
            Pichler, Hannes and
            others},
  title = {Parallel Implementation of High-Fidelity Multiqubit Gates with Neutral Atoms},
  journal = {Phys. Rev. Lett.},
  volume = {123},
  number = {17},
  pages = {170503},
  publisher = {American Physical Society},
  year = {2019},
  doi = {10.1103/PhysRevLett.123.170503}
}

@article{PRXQuantum.6.010331,
  author = {Tsai, Richard Bing-Shiun and
            Sun, Xiangkai and
            Shaw, Adam L. and
            Finkelstein, Ran and
            Endres, Manuel},
  title = {Benchmarking and Fidelity Response Theory of High-Fidelity {Rydberg} Entangling Gates},
  journal = {PRX Quantum},
  volume = {6},
  number = {1},
  pages = {010331},
  publisher = {American Physical Society},
  year = {2025},
  doi = {10.1103/PRXQuantum.6.010331}
}

@article{PRXQuantum.6.020334,
  author = {Muniz, J. A. and
            Stone, M. and
            Stack, D. T. and
            Jaffe, M. and
            Kindem, J. M. and
            Wadleigh, L. and
            Zalys-Geller, E. and
            Zhang, X. and
            Chen, C.-A. and
            Norcia, M. A. and
            others},
  title = {High-Fidelity Universal Gates in the $^{171}\mathrm{Yb}$ Ground-State Nuclear-Spin Qubit},
  journal = {PRX Quantum},
  volume = {6},
  number = {2},
  pages = {020334},
  publisher = {American Physical Society},
  year = {2025},
  doi = {10.1103/PRXQuantum.6.020334}
}

@article{2025saffman,
  author = {Radnaev, A.G. and
            Chung, W.C. and
            Cole, D.C. and
            Mason, D. and
            Ballance, T.G. and
            Bedalov, M.J. and
            Belknap, D.A. and
            Berman, M.R. and
            Blakely, M. and
            Bloomfield, I.L. and
            others},
  title = {Universal Neutral-Atom Quantum Computer with Individual Optical Addressing and Nondestructive Readout},
  journal = {PRX Quantum},
  volume = {6},
  number = {3},
  pages = {030334},
  publisher = {American Physical Society},
  year = {2025},
  doi = {10.1103/66s8-jj18}
}

@article{lukin2026highfidelity,
  author  = {Evered, Simon J. and
            Xu, Muqing and
            Li, Sophie H. and
            Geim, Alexandra A. and
            Bonilla Ataides, J. Pablo and
            Kalinowski, Marcin and
            Bluvstein, Dolev and
            Maskara, Nishad and
            Kokail, Christian and
            Greiner, Markus and
            others},
  title   = {High-fidelity entangling gates and nonlocal circuits with neutral atoms},
  journal = {Science},
  year    = {2026},
  volume  = {394},
  number  = {6820},
  pages   = {197--201},
  doi     = {10.1126/science.aei1312}
}

@article{Graham2022NeutralAtomAlgorithms,
  author = {Graham, T. M. and
            Song, Y. and
            Scott, J. and
            Poole, C. and
            Phuttitarn, L. and
            Jooya, K. and
            Eichler, P. and
            Jiang, X. and
            Marra, A. and
            Grinkemeyer, B. and
            others},
  title = {Multi-qubit entanglement and algorithms on a neutral-atom quantum computer},
  journal = {Nature},
  volume = {604},
  number = {7906},
  pages = {457--462},
  year = {2022},
  doi = {10.1038/s41586-022-04603-6}
}

@article{Norcia2023,
  author = {Norcia, M. A. and
            Cairncross, W. B. and
            Barnes, K. and
            Battaglino, P. and
            Brown, A. and
            Brown, M. O. and
            Cassella, K. and
            Chen, C.-A. and
            Coxe, R. and
            Crow, D. and
            others},
  title = {Midcircuit Qubit Measurement and Rearrangement in a $^{171}\mathrm{Yb}$ Atomic Array},
  journal = {Phys. Rev. X},
  volume = {13},
  number = {4},
  pages = {041034},
  publisher = {American Physical Society},
  year = {2023},
  doi = {10.1103/PhysRevX.13.041034}
}

@article{2023science,
  author = {K. Singh and
            C. E. Bradley and
            S. Anand and
            V. Ramesh and
            R. White and
            H. Bernien},
  title = {Mid-circuit correction of correlated phase errors using an array of spectator qubits},
  journal = {Science},
  volume = {380},
  number = {6651},
  pages = {1265--1269},
  year = {2023},
  doi = {10.1126/science.ade5337}
}

@article{Bluvstein2024,
  author = {Bluvstein, Dolev and
            Evered, Simon J. and
            Geim, Alexandra A. and
            Li, Sophie H. and
            Zhou, Hengyun and
            Manovitz, Tom and
            Ebadi, Sepehr and
            Cain, Madelyn and
            Kalinowski, Marcin and
            Hangleiter, Dominik and
            others},
  title = {Logical quantum processor based on reconfigurable atom arrays},
  journal = {Nature},
  volume = {626},
  number = {7997},
  pages = {58--65},
  year = {2024},
  doi = {10.1038/s41586-023-06927-3}
}

@article{Bluvstein2026,
  author = {Bluvstein, Dolev and
            Geim, Alexandra A. and
            Li, Sophie H. and
            Evered, Simon J. and
            Bonilla Ataides, J. Pablo and
            Baranes, Gefen and
            Gu, Andi and
            Manovitz, Tom and
            Xu, Muqing and
            Kalinowski, Marcin and
            others},
  title = {A fault-tolerant neutral-atom architecture for universal quantum computation},
  journal = {Nature},
  volume = {649},
  number = {8095},
  pages = {39--46},
  year = {2026},
  doi = {10.1038/s41586-025-09848-5}
}

@article{address2015,
  author = {Xia, T. and
            Lichtman, M. and
            Maller, K. and
            Carr, A. W. and
            Piotrowicz, M. J. and
            Isenhower, L. and
            Saffman, M.},
  title = {Randomized Benchmarking of Single-Qubit Gates in a {2D} Array of Neutral-Atom Qubits},
  journal = {Phys. Rev. Lett.},
  volume = {114},
  number = {10},
  pages = {100503},
  publisher = {American Physical Society},
  year = {2015},
  doi = {10.1103/PhysRevLett.114.100503}
}

@article{Graham2023MultiscaleAddressing,
  author = {Graham, T. M. and
            Oh, E. and
            Saffman, M.},
  title = {Multiscale architecture for fast optical addressing and control of large-scale qubit arrays},
  journal = {Appl. Opt.},
  volume = {62},
  number = {12},
  pages = {3242--3251},
  year = {2023},
  doi = {10.1364/AO.484367}
}

@article{Zhang2024ScaledLocalControl,
  author = {Zhang, Bichen and
            Peng, Pai and
            Paul, Aditya and
            Thompson, Jeff D.},
  title = {Scaled local gate controller for optically addressed qubits},
  journal = {Optica},
  volume = {11},
  number = {2},
  pages = {227--233},
  year = {2024},
  doi = {10.1364/OPTICA.512155}
}

@article{Menssen2023PhotonicControl,
  author = {Menssen, Adrian J. and
            Hermans, Artur and
            Christen, Ian and
            Propson, Thomas and
            Li, Chao and
            Leenheer, Andrew J. and
            Zimmermann, Matthew and
            Dong, Mark and
            Larocque, Hugo and
            Raniwala, Hamza and
            others},
  title = {Scalable photonic integrated circuits for high-fidelity light control},
  journal = {Optica},
  volume = {10},
  number = {10},
  pages = {1366--1372},
  year = {2023},
  doi = {10.1364/OPTICA.489504}
}

@article{rbcs2022,
  author = {Singh, Kevin and
            Anand, Shraddha and
            Pocklington, Andrew and
            Kemp, Jordan T. and
            Bernien, Hannes},
  title = {Dual-Element, Two-Dimensional Atom Array with Continuous-Mode Operation},
  journal = {Phys. Rev. X},
  volume = {12},
  number = {1},
  pages = {011040},
  publisher = {American Physical Society},
  year = {2022},
  doi = {10.1103/PhysRevX.12.011040}
}

@article{Beterov2015DifferentSpecies,
  author = {Beterov, I. I. and
            Saffman, M.},
  title = {{Rydberg} blockade, {F{\"o}rster} resonances, and quantum state measurements with different atomic species},
  journal = {Phys. Rev. A},
  volume = {92},
  number = {4},
  pages = {042710},
  year = {2015},
  doi = {10.1103/PhysRevA.92.042710}
}

@article{Petrosyan2024DualSpeciesGates,
  author = {Petrosyan, D. and
            Norrell, S. and
            Poole, C. and
            Saffman, M.},
  title = {Fast measurements and multiqubit gates in dual-species atomic arrays},
  journal = {Phys. Rev. A},
  volume = {110},
  number = {4},
  pages = {042404},
  year = {2024},
  doi = {10.1103/PhysRevA.110.042404}
}

@article{2024np,
  author = {Anand, Shraddha and
            Bradley, Conor E. and
            White, Ryan and
            Ramesh, Vikram and
            Singh, Kevin and
            Bernien, Hannes},
  title = {A dual-species {Rydberg} array},
  journal = {Nat. Phys.},
  volume = {20},
  number = {11},
  pages = {1744--1750},
  year = {2024},
  doi = {10.1038/s41567-024-02638-2}
}

@article{2026saffman,
  author  = {Miles, J. and
            Lichtman, M. T. and
            Scott, A. M. and
            Scott, J. and
            Norrell, S. A. and
            Bedalov, M. J. and
            Belknap, D. A. and
            Cole, D. C. and
            Eubanks, S. Y. and
            Gillette, M. and
            others},
  title   = {Qubit Syndrome Measurements with a High-Fidelity {Rb}-{Cs} {Rydberg} Gate},
  journal = {Phys. Rev. Lett.},
  year    = {2026},
  volume  = {137},
  number  = {10},
  pages   = {100601},
  doi     = {10.1103/pvjm-jxd2}
}

@misc{2026qca,
  author = {Ryan White and
            Vikram Ramesh and
            Alexander Impertro and
            Shraddha Anand and
            Francesco Cesa and
            Giuliano Giudici and
            Thomas Iadecola and
            Hannes Pichler and
            Hannes Bernien},
  title = {Quantum Cellular Automata on a Dual-Species {Rydberg} Processor},
  year = {2026},
  eprint = {2601.16257},
  archiveprefix = {arXiv},
  primaryclass = {quant-ph},
  url = {https://arxiv.org/abs/2601.16257}
}

@article{fidelity2021,
  author = {Kliesch, Martin and
            Roth, Ingo},
  title = {Theory of Quantum System Certification},
  journal = {PRX Quantum},
  volume = {2},
  number = {1},
  pages = {010201},
  publisher = {American Physical Society},
  year = {2021},
  doi = {10.1103/PRXQuantum.2.010201}
}

@article{xzzx2021,
  author = {Bonilla Ataides, J. Pablo and
            Tuckett, David K. and
            Bartlett, Stephen D. and
            Flammia, Steven T. and
            Brown, Benjamin J.},
  title = {The {XZZX} surface code},
  journal = {Nat. Commun.},
  volume = {12},
  number = {1},
  pages = {2172},
  year = {2021},
  doi = {10.1038/s41467-021-22274-1}
}

@article{zxxz,
  author = {Gidney, Craig},
  title = {Inplace Access to the Surface Code {Y} Basis},
  journal = {Quantum},
  volume = {8},
  pages = {1310},
  publisher = {Verein zur Forderung des Open Access Publizierens in den Quantenwissenschaften},
  year = {2024},
  doi = {10.22331/q-2024-04-08-1310}
}

@article{Google2023,
  author = {{Google Quantum AI}},
  title = {Suppressing quantum errors by scaling a surface code logical qubit},
  journal = {Nature},
  volume = {614},
  number = {7949},
  pages = {676--681},
  year = {2023},
  doi = {10.1038/s41586-022-05434-1}
}

@article{Google2025,
  author = {{Google Quantum AI and Collaborators}},
  title = {Quantum error correction below the surface code threshold},
  journal = {Nature},
  volume = {638},
  number = {8052},
  pages = {920--926},
  year = {2025},
  doi = {10.1038/s41586-024-08449-y}
}

@article{pymatching,
  author = {Higgott, Oscar and
            Gidney, Craig},
  title = {Sparse {Blossom}: correcting a million errors per core second with minimum-weight matching},
  journal = {Quantum},
  volume = {9},
  pages = {1600},
  year = {2025},
  doi = {10.22331/q-2025-01-20-1600}
}

@article{logicals2026,
  author = {Chen, Zi-Han and
            Chen, Ming-Cheng and
            Lu, Chao-Yang and
            Pan, Jian-Wei},
  title = {Transversal Logical {Clifford} Gates on the Rotated Surface Code with Reconfigurable Neutral Atom Arrays},
  journal = {Phys. Rev. Lett.},
  volume = {136},
  number = {13},
  pages = {130601},
  publisher = {American Physical Society},
  year = {2026},
  doi = {10.1103/m7tq-9v3g}
}

@article{Breuckmann2024foldtransversal,
  author = {Breuckmann, Nikolas P. and
            Burton, Simon},
  title = {Fold-transversal {Clifford} gates for quantum codes},
  journal = {{Quantum}},
  volume = {8},
  pages = {1372},
  publisher = {{Verein zur F{\"{o}}rderung des Open Access Publizierens in den Quantenwissenschaften}},
  year = {2024},
  doi = {10.22331/q-2024-06-13-1372}
}

@misc{2024msc,
  author = {Craig Gidney and
            Noah Shutty and
            Cody Jones},
  title = {Magic state cultivation: growing {T} states as cheap as {CNOT} gates},
  year = {2024},
  eprint = {2409.17595},
  archiveprefix = {arXiv},
  primaryclass = {quant-ph},
  url = {https://arxiv.org/abs/2409.17595}
}

@article{sahay2026foldmsc,
  author = {Kaavya Sahay and
            Pei-Kai Tsai and
            Kathleen Chang and
            Qile Su and
            Thomas B. Smith and
            Shraddha Singh and
            Shruti Puri},
  title = {Fold-transversal surface code cultivation},
  journal = {PRX Quantum},
  volume = {7},
  pages = {033006},
  year = {2026},
  doi = {10.1103/gpvl-lg4c}
}

@article{p8tw-6kq9,
  author = {Vaknin, Yotam and
            Jacoby, Shoham and
            Grimsmo, Arne and
            Retzker, Alex},
  title = {High Rate Magic State Cultivation on the Surface Code},
  journal = {PRX Quantum},
  volume = {7},
  number = {1},
  pages = {010353},
  publisher = {American Physical Society},
  year = {2026},
  doi = {10.1103/p8tw-6kq9}
}

@misc{claes2025msc,
  author = {Jahan Claes},
  title = {Cultivating {T} states on the surface code with only two-qubit gates},
  year = {2025},
  eprint = {2509.05232},
  archiveprefix = {arXiv},
  primaryclass = {quant-ph},
  url = {https://arxiv.org/abs/2509.05232}
}

@article{Piroli2021LOCC,
  author = {Piroli, Lorenzo and
            Styliaris, Georgios and
            Cirac, J. Ignacio},
  title = {Quantum Circuits Assisted by Local Operations and Classical Communication: Transformations and Phases of Matter},
  journal = {Phys. Rev. Lett.},
  volume = {127},
  number = {22},
  pages = {220503},
  publisher = {American Physical Society},
  year = {2021},
  doi = {10.1103/PhysRevLett.127.220503}
}

@article{Xu2024,
  author = {Xu, Qian and
            Bonilla Ataides, J. Pablo and
            Pattison, Christopher A. and
            Raveendran, Nithin and
            Bluvstein, Dolev and
            Wurtz, Jonathan and
            Vasi{\'{c}}, Bane and
            Lukin, Mikhail D. and
            Jiang, Liang and
            Zhou, Hengyun},
  title = {Constant-overhead fault-tolerant quantum computation with reconfigurable atom arrays},
  journal = {Nat. Phys.},
  volume = {20},
  number = {7},
  pages = {1084--1090},
  year = {2024},
  doi = {10.1038/s41567-024-02479-z}
}

@article{fidelity1999,
  author = {Horodecki, Micha\l{} and
            Horodecki, Pawe\l{} and
            Horodecki, Ryszard},
  title = {General teleportation channel, singlet fraction, and quasidistillation},
  journal = {Phys. Rev. A},
  volume = {60},
  number = {3},
  pages = {1888--1898},
  publisher = {American Physical Society},
  year = {1999},
  doi = {10.1103/PhysRevA.60.1888}
}

@article{Nielsen2002,
  author = {Nielsen, Michael A},
  title = {A simple formula for the average gate fidelity of a quantum dynamical operation},
  journal = {Phys. Lett. A},
  volume = {303},
  number = {4},
  pages = {249--252},
  publisher = {Elsevier BV},
  year = {2002},
  doi = {10.1016/s0375-9601(02)01272-0}
}

@article{Fowler2012SurfaceCodes,
  author = {Fowler, Austin G. and
            Mariantoni, Matteo and
            Martinis, John M. and
            Cleland, Andrew N.},
  title = {Surface codes: Towards practical large-scale quantum computation},
  journal = {Phys. Rev. A},
  volume = {86},
  number = {3},
  pages = {032324},
  year = {2012},
  doi = {10.1103/PhysRevA.86.032324}
}

@article{Stephens2014SurfaceCodeThresholds,
  author = {Stephens, Ashley M.},
  title = {Fault-tolerant thresholds for quantum error correction with the surface code},
  journal = {Phys. Rev. A},
  volume = {89},
  number = {2},
  pages = {022321},
  year = {2014},
  doi = {10.1103/PhysRevA.89.022321}
}

@article{PhysRevLett.131.170601,
  author = {Cesa, Francesco and
            Pichler, Hannes},
  title = {Universal Quantum Computation in Globally Driven {Rydberg} Atom Arrays},
  journal = {Phys. Rev. Lett.},
  volume = {131},
  number = {17},
  pages = {170601},
  year = {2023},
  doi = {10.1103/PhysRevLett.131.170601}
}

@misc{Roffe_LDPC_Python_tools_2022,
  author = {Roffe, Joschka},
  title = {{LDPC}: {Python} tools for low-density parity-check codes},
  year = {2022},
  url = {https://pypi.org/project/ldpc/}
}

@article{roffe_decoding_2020,
  author = {Roffe, Joschka and
            White, David R. and
            Burton, Simon and
            Campbell, Earl},
  title = {Decoding across the quantum low-density parity-check code landscape},
  journal = {Phys. Rev. Research},
  volume = {2},
  number = {4},
  pages = {043423},
  year = {2020},
  doi = {10.1103/PhysRevResearch.2.043423}
}

@article{Beugnon2007Transport,
  author = {Beugnon, J{\'e}r{\^o}me and
            Tuchendler, Charles and
            Marion, Harold and
            Ga{\"e}tan, Alpha and
            Miroshnychenko, Yevhen and
            Sortais, Yvan R. P. and
            Lance, Andrew M. and
            Jones, Matthew P. A. and
            Messin, Ga{\'e}tan and
            Browaeys, Antoine and
            others},
  title = {Two-dimensional transport and transfer of a single atomic qubit in optical tweezers},
  journal = {Nat. Phys.},
  volume = {3},
  number = {10},
  pages = {696--699},
  year = {2007},
  doi = {10.1038/nphys698}
}

@misc{Zeng2026FastReadout,
  author = {Zeng, Xu-Zhao-Qiu and
            You, Chang and
            Wang, Qing-Wei and
            Li, Zi-Feng and
            Ji, Yi and
            An, Dong and
            Yu, Chao and
            Liu, Jia-Rui and
            He, Zi-Mo and
            Gu, Jia-Rui and
            others},
  title = {Fast nondestructive readout for high-clock-rate atom array quantum processor},
  year = {2026},
  eprint = {2608.17189},
  archiveprefix = {arXiv},
  primaryclass = {quant-ph},
  url = {https://arxiv.org/abs/2608.17189}
}

@article{Gidney2021RSA2048,
  author  = {Gidney, Craig and
            Eker{\r{a}}, Martin},
  title   = {How to factor 2048 bit {RSA} integers in 8 hours using 20 million noisy qubits},
  journal = {Quantum},
  year    = {2021},
  volume  = {5},
  pages   = {433},
  doi     = {10.22331/q-2021-04-15-433}
}

@article{Reiher2017ReactionMechanisms,
  author  = {Reiher, Markus and
            Wiebe, Nathan and
            Svore, Krysta M. and
            Wecker, Dave and
            Troyer, Matthias},
  title   = {Elucidating reaction mechanisms on quantum computers},
  journal = {Proc. Natl Acad. Sci. USA},
  year    = {2017},
  volume  = {114},
  number  = {29},
  pages   = {7555--7560},
  doi     = {10.1073/pnas.1619152114}
}

@article{Low2025SpectralAmplification,
  author  = {Low, Guang Hao and
            King, Robbie and
            Berry, Dominic W. and
            Han, Qiushi and
            DePrince, III, A. Eugene and
            White, Alec F. and
            Babbush, Ryan and
            Somma, Rolando D. and
            Rubin, Nicholas C.},
  title   = {Fast Quantum Simulation of Electronic Structure by Spectral Amplification},
  journal = {Phys. Rev. X},
  year    = {2025},
  volume  = {15},
  number  = {4},
  pages   = {041016},
  doi     = {10.1103/pb2g-j9cw}
}

@article{Zhou2025TransversalResources,
  author    = {Zhou, Hengyun and
            Duckering, Casey and
            Zhao, Chen and
            Bluvstein, Dolev and
            Cain, Madelyn and
            Kubica, Aleksander and
            Wang, Sheng-Tao and
            Lukin, Mikhail D.},
  title     = {Resource Analysis of Low-Overhead Transversal Architectures for Reconfigurable Atom Arrays},
  booktitle = {Proceedings of the 52nd Annual International Symposium on Computer Architecture},
  journal   = {Proceedings of the 52nd Annual International Symposium on Computer Architecture},
  year      = {2025},
  pages     = {1432--1448},
  publisher = {Association for Computing Machinery},
  doi       = {10.1145/3695053.3731039}
}

@misc{Cain2026ShorAtomicQubits,
  author       = {Cain, Madelyn and
            Xu, Qian and
            King, Robbie and
            Picard, Lewis R. B. and
            Levine, Harry and
            Endres, Manuel and
            Preskill, John and
            Huang, Hsin-Yuan and
            Bluvstein, Dolev},
  title        = {{Shor}'s algorithm is possible with as few as 10,000 reconfigurable atomic qubits},
  year         = {2026},
  howpublished = {Preprint at \url{https://arxiv.org/abs/2603.28627}},
  doi          = {10.48550/arXiv.2603.28627}
}

@article{Weimer2010RydbergSimulator,
  author  = {Weimer, Hendrik and
            M{\"{u}}ller, Markus and
            Lesanovsky, Igor and
            Zoller, Peter and
            B{\"{u}}chler, Hans Peter},
  title   = {A {Rydberg} quantum simulator},
  journal = {Nat. Phys.},
  year    = {2010},
  volume  = {6},
  number  = {5},
  pages   = {382--388},
  doi     = {10.1038/nphys1614}
}

@article{Chen2026RPCultivation,
  author  = {Chen, Zi-Han and
            Chen, Ming-Cheng and
            Lu, Chao-Yang and
            Pan, Jian-Wei},
  title   = {Efficient magic state cultivation on $\mathbb{RP}^{2}$},
  journal = {PRX Quantum},
  year    = {2026},
  volume  = {7},
  number  = {1},
  pages   = {010315},
  doi     = {10.1103/9kys-3whh}
}

@article{Sarkany2015MicrowaveCavity,
  author  = {S{\'{a}}rk{\'{a}}ny, L{\H{o}}rinc and
            Fort{\'{a}}gh, J{\'{o}}zsef and
            Petrosyan, David},
  title   = {Long-range quantum gate via {Rydberg} states of atoms in a thermal microwave cavity},
  journal = {Phys. Rev. A},
  year    = {2015},
  volume  = {92},
  number  = {3},
  pages   = {030303},
  doi     = {10.1103/physreva.92.030303}
}

@article{Sheng2018Microwave,
  author = {Sheng, Cheng and
            He, Xiaodong and
            Xu, Peng and
            Guo, Ruijun and
            Wang, Kunpeng and
            Xiong, Zongyuan and
            Liu, Min and
            Wang, Jin and
            Zhan, Mingsheng},
  title = {High-Fidelity Single-Qubit Gates on Neutral Atoms in a
           Two-Dimensional Magic-Intensity Optical Dipole Trap Array},
  journal = {Phys. Rev. Lett.},
  volume = {121},
  number = {24},
  pages = {240501},
  year = {2018},
  doi = {10.1103/PhysRevLett.121.240501}
}

@article{Nikolov2023Microwave,
  author = {Nikolov, B. and
            Diamond-Hitchcock, E. and
            Bass, J. and
            Spong, N. L. R. and
            Pritchard, J. D.},
  title = {Randomized Benchmarking Using Nondestructive Readout
           in a Two-Dimensional Atom Array},
  journal = {Phys. Rev. Lett.},
  volume = {131},
  number = {3},
  pages = {030602},
  year = {2023},
  doi = {10.1103/PhysRevLett.131.030602}
}

@article{WoodGambetta2018,
  author  = {Wood, Christopher J. and Gambetta, Jay M.},
  title   = {Quantification and characterization of leakage errors},
  journal = {Physical Review A},
  volume  = {97},
  number  = {3},
  pages   = {032306},
  year    = {2018},
  doi     = {10.1103/PhysRevA.97.032306}
}

@article{Suchara2015Leakage,
  author  = {Suchara, Martin and Cross, Andrew W. and Gambetta, Jay M.},
  title   = {Leakage suppression in the toric code},
  journal = {Quantum Information and Computation},
  volume  = {15},
  number  = {11--12},
  pages   = {997--1016},
  year    = {2015},
  doi     = {10.26421/QIC15.11-12-8}
}

@article{Bombin2024FTPostselection,
  author  = {Bomb{\'i}n, H{\'e}ctor and
             Pant, Mihir and
             Roberts, Sam and
             Seetharam, Karthik I.},
  title   = {Fault-tolerant postselection for low-overhead magic state preparation},
  journal = {PRX Quantum},
  volume  = {5},
  pages   = {010302},
  year    = {2024},
  doi     = {10.1103/PRXQuantum.5.010302}
}

@article{Zhang2024CavityDarkMode,
  author = {Zhang, Xiaotian and Yu, Zhanhai and
            Zhang, Hongrui and Xiang, Di and Zhang, Hao},
  title = {Cavity dark mode mediated by atom array
           without atomic scattering loss},
  journal = {Phys. Rev. Research},
  volume = {6},
  pages = {L042026},
  year = {2024},
  doi = {10.1103/PhysRevResearch.6.L042026}
}

@article{Jandura2022TimeOptimal,
  author  = {Jandura, Sven and Pupillo, Guido},
  title   = {Time-optimal two- and three-qubit gates for {Rydberg} atoms},
  journal = {Quantum},
  volume  = {6},
  pages   = {712},
  year    = {2022},
  doi     = {10.22331/q-2022-05-13-712}
}

@article{Pagano2022ErrorBudgeting,
  author  = {Pagano, Alice and
             Weber, Sebastian and
             Jaschke, Daniel and
             Pfau, Tilman and
             Meinert, Florian and
             Montangero, Simone and
             B{\"u}chler, Hans Peter},
  title   = {Error budgeting for a controlled-phase gate with strontium-88 {Rydberg} atoms},
  journal = {Phys. Rev. Res.},
  volume  = {4},
  pages   = {033019},
  year    = {2022},
  doi     = {10.1103/PhysRevResearch.4.033019}
}

@misc{Bhardwaj2026HighRateQLDPC,
  author       = {Bhardwaj, Aditya and Ma, Muzhou and Meister, Nadine and
                  King, Robbie and Bluvstein, Dolev and Preskill, John and
                  Cain, Madelyn and Xu, Qian and Huang, Hsin-Yuan},
  title        = {High-rate {qLDPC} processors},
  howpublished = {Preprint at \url{https://arxiv.org/abs/2607.28795}},
  year         = {2026},
  doi          = {10.48550/arXiv.2607.28795}
}

@misc{Chen2026MicrowaveCatBus,
  author       = {Chen, Yanyan and Yu, Xinyang and Min, Yueyang and
                  Zhang, Zhihao and Cao, Shuaifan and Li, Xiaopeng},
  title        = {Fault-tolerant quantum computing with a microwave {Cat Bus}},
  howpublished = {Preprint at \url{https://arxiv.org/abs/2608.01111}},
  year         = {2026},
  doi          = {10.48550/arXiv.2608.01111}
}

@article{Berthusen2025LocalQLDPC,
  author  = {Berthusen, Noah and Devulapalli, Dhruv and Schoute, Eddie and
             Childs, Andrew M. and Gullans, Michael J. and Gorshkov, Alexey V. and
             Gottesman, Daniel},
  title   = {Toward a {2D} local implementation of quantum low-density parity-check codes},
  journal = {PRX Quantum},
  year    = {2025},
  volume  = {6},
  pages   = {010306},
  doi     = {10.1103/PRXQuantum.6.010306}
}

@misc{Zhao2026UltraHighRate,
  author       = {Zhao, Chen and Duckering, Casey and Gu, Andi and Maskara, Nishad and Zhou, Hengyun},
  title        = {Towards ultra-high-rate quantum error correction with reconfigurable atom arrays},
  howpublished = {Preprint at \url{https://arxiv.org/abs/2604.16209}},
  year         = {2026},
  doi          = {10.48550/arXiv.2604.16209}
}

@article{Bergonzoni2026StaticBuses,
  author  = {Bergonzoni, Matteo and Pecorari, Laura and Norrell, Sam and
             Poole, Cody and Pupillo, Guido and Saffman, Mark},
  title   = {Fault-tolerant quantum computation with static atomic buses},
  journal = {PRX Quantum},
  year    = {2026},
  note    = {Accepted for publication},
  eprint  = {2607.02804},
  archiveprefix = {arXiv},
  doi     = {10.1103/n1ll-yxtt}
}

@misc{Shaw2026Networked,
  author       = {Shaw, Swayangprabha and Rengaswamy, Narayanan},
  title        = {Networked realization of quantum {LDPC} codes},
  howpublished = {Preprint at \url{https://arxiv.org/abs/2604.25026}},
  year         = {2026},
  doi          = {10.48550/arXiv.2604.25026}
}

@misc{Pecorari2025Addressable,
  author       = {Pecorari, Laura and Guerci, Francesco Paolo and Perrin, Hugo and Pupillo, Guido},
  title        = {Addressable gate-based logical computation with quantum {LDPC} codes},
  howpublished = {Preprint at \url{https://arxiv.org/abs/2511.06124}},
  year         = {2025},
  doi          = {10.48550/arXiv.2511.06124}
}

@misc{Khan2026EarlyFT,
  author       = {Khan, Sahil and Sethi, Sayam and Sahay, Kaavya and Lin, Yingjia and
                  Alnas, Jude and Kurapati, Suhas and Anand, Abhinav and Baker, Jonathan M. and Brown, Kenneth R.},
  title        = {Architecting early fault tolerant neutral atoms systems with quantum advantage},
  howpublished = {Preprint at \url{https://arxiv.org/abs/2604.19735}},
  year         = {2026},
  doi          = {10.48550/arXiv.2604.19735}
}

@misc{Zhou2026AlgebraicSurgery,
  author       = {Zhou, Tian-Gang and Gu, Boren and Eisert, Jens and Zhao, Chen},
  title        = {Parallel algebraic surgery for {qLDPC} codes with constant shuttling depth},
  howpublished = {Preprint at \url{https://arxiv.org/abs/2609.39874}},
  year         = {2026},
  doi          = {10.48550/arXiv.2609.39874}
}

@article{Stack2026Transversal,
  author  = {Stack, John and Wang, Ming and Mueller, Frank},
  title   = {Transversal fault tolerant distributed quantum computing operations},
  journal = {Nat. Commun.},
  year    = {2026},
  volume  = {17},
  pages   = {8838},
  doi     = {10.1038/s41467-026-75693-3}
}

@article{Viszlai2026QSIEVE,
  author  = {Viszlai, Joshua and Yang, Willers and Lin, Sophia and Liu, Junyu and
             Nottingham, Natalia and Baker, Jonathan M. and Chong, Frederic T.},
  title   = {{qSIEVE}: Efficient {qLDPC} memory via systolic movement in atom arrays},
  journal = {ACM Trans. Quantum Comput.},
  year    = {2026},
  volume  = {7},
  number  = {2},
  articleno = {9},
  doi     = {10.1145/3779066}
}

@misc{Gidney2017Factoring,
  author       = {Gidney, Craig},
  title        = {Factoring with {$n+2$} clean qubits and {$n-1$} dirty qubits},
  howpublished = {Preprint at \url{https://arxiv.org/abs/1706.07884}},
  year         = {2017},
  doi          = {10.48550/arXiv.1706.07884}
}

@incollection{Haner2020EllipticCurve,
  author    = {H{\"a}ner, Thomas and Jaques, Samuel and Naehrig, Michael and
               Roetteler, Martin and Soeken, Mathias},
  title     = {Improved quantum circuits for elliptic curve discrete logarithms},
  editor    = {Ding, Jintai and Tillich, Jean-Pierre},
  booktitle = {Post-Quantum Cryptography},
  series    = {Lecture Notes in Computer Science},
  volume    = {12100},
  pages     = {425--444},
  publisher = {Springer International Publishing},
  address   = {Cham},
  year      = {2020},
  doi       = {10.1007/978-3-030-44223-1_23}
}

@article{Babbush2018ElectronicSpectra,
  author  = {Babbush, Ryan and Gidney, Craig and Berry, Dominic W. and
             Wiebe, Nathan and McClean, Jarrod and Paler, Alexandru and
             Fowler, Austin and Neven, Hartmut},
  title   = {Encoding electronic spectra in quantum circuits with linear {T} complexity},
  journal = {Phys. Rev. X},
  volume  = {8},
  pages   = {041015},
  year    = {2018},
  doi     = {10.1103/PhysRevX.8.041015}
}

@misc{Gidney2025RSA2048,
  author       = {Gidney, Craig},
  title        = {How to factor 2048 bit {RSA} integers with less than a million noisy qubits},
  howpublished = {Preprint at \url{https://arxiv.org/abs/2505.15917}},
  year         = {2025},
  doi          = {10.48550/arXiv.2505.15917}
}

@article{Zhou2025FullStackDesign,
  author  = {Zhou, Hengyun and Cain, Madelyn and Lukin, Mikhail D.},
  title   = {Opportunities in full-stack design of low-overhead fault-tolerant quantum computation},
  journal = {Nat. Comput. Sci.},
  volume  = {5},
  number  = {12},
  pages   = {1110--1119},
  year    = {2025},
  doi     = {10.1038/s43588-025-00895-6}
}

@article{Gidney2021faulttolerant,
  author = {Gidney, Craig and Newman, Michael and Fowler, Austin and Broughton, Michael},
  title = {A fault-tolerant honeycomb memory},
  journal = {Quantum},
  volume = {5},
  pages = {605},
  year = {2021},
  doi = {10.22331/q-2021-12-20-605}
}

@misc{ORourke2024ComparePair,
  author       = {O'Rourke, Anthony Ryan and Devitt, Simon},
  title        = {Compare the Pair: Rotated vs. Unrotated Surface Codes at Equal Logical Error Rates},
  howpublished = {Preprint at \url{https://arxiv.org/abs/2409.14765}},
  year         = {2024},
  doi          = {10.48550/arXiv.2409.14765}
}

@article{Powell1964,
  author  = {Powell, M. J. D.},
  title   = {An efficient method for finding the minimum of a function
             of several variables without calculating derivatives},
  journal = {The Computer Journal},
  volume  = {7},
  number  = {2},
  pages   = {155--162},
  year    = {1964},
  doi     = {10.1093/comjnl/7.2.155}
}

@article{Gidney2025Yoked,
  author  = {Gidney, Craig and Newman, Michael and Brooks, Peter and Jones, Cody},
  title   = {Yoked surface codes},
  journal = {Nature Communications},
  volume  = {16},
  pages   = {4498},
  year    = {2025},
  doi     = {10.1038/s41467-025-59714-1}
}

@article{Hutter2014SurfaceCode,
  author  = {Hutter, Adrian and Wootton, James R. and Loss, Daniel},
  title   = {Efficient {Markov} chain {Monte Carlo} algorithm for the surface code},
  journal = {Physical Review A},
  volume  = {89},
  pages   = {022326},
  year    = {2014},
  doi     = {10.1103/PhysRevA.89.022326}
}

@misc{Meister2024SoftOutput,
  author       = {Meister, Nadine and Pattison, Christopher A. and Preskill, John},
  title        = {Efficient soft-output decoders for the surface code},
  howpublished = {Preprint at \url{https://arxiv.org/abs/2405.07433}},
  year         = {2024}
}

\clearpage
\section*{Methods}
\label{sec:methods}

\begingroup
\small
\let\ncsMethodsSubsection\subsection
\let\ncsMethodsSubsubsection\subsubsection
\let\ncsMethodsParagraph\paragraph
\renewcommand{\section}[1]{\ncsMethodsSubsection*{#1}}
\renewcommand{\subsection}[1]{\ncsMethodsSubsubsection*{#1}}
\renewcommand{\subsubsection}[1]{\ncsMethodsParagraph*{#1}}

\makeatletter
\renewcommand*{\theHequation}{methods.\arabic{section}.\arabic{equation}}
\makeatother

\section{Hardware-Inferred Circuit-Level Noise Model}
\label{app:heterogeneous_noise}

We derive a hardware-inferred circuit-level noise model from \(\mathsf{GRAIL}\) by combining numerical simulations of Rydberg pulses $\mathrm{Ryd}_s(\varphi)$ with effective error channels for global microwave rotation $G_s(U)$, reset $\mathcal R_s$ and measurement $\mathcal M_s$. 
Our noise model for global Rydberg pulses accounts for spontaneous Rydberg decay and finite blockade strength, which are expected to be intrinsic limitations of Rydberg gates~\cite{Jandura2022TimeOptimal,Pagano2022ErrorBudgeting}.
We calculate the resulting error channels for participating atom pairs and idling atoms separately, for each global Rydberg pulse at each Rabi frequency \(\Omega\).
The resulting channels are incorporated into the compiled \(\mathsf{GRAIL}\) circuits to evaluate fault-tolerant performance.

During the global Rydberg pulse to implement $\rm CP(\varphi)$, the Rb and Cs atoms are driven with the same Rabi frequency \(\Omega\) and phase profile~\cite{Jandura2022TimeOptimal,highfidelity}.
For a participating atom pair, the driven Hamiltonian is
\[
H(t)=
\frac{\Omega}{2}
\left[
e^{-i\phi(t)}
\sum_{j=1,2}
\lvert r_j\rangle\langle1_j\rvert
+\mathrm{h.c.}
\right]
+V\lvert r_1r_2\rangle\langle r_1r_2\rvert .
\]
We simulate the evolution of the pair by solving the Lindblad master equation
\begin{equation}
\frac{\mathrm d\rho}{\mathrm dt}
=-i[H(t),\rho]
+\sum_{j=1,2}\Gamma_j
\left(
L_j\rho L_j^\dagger
-\frac{1}{2}\{L_j^\dagger L_j,\rho\}
\right),
\label{eq:rydberg_lindblad}
\end{equation}
where \(L_j=\lvert e_j\rangle\langle r_j\rvert\) describes spontaneous Rydberg decay at rate \(\Gamma_j\). 
The state \(\lvert e_j\rangle\) represents an effective error state outside the computational subspace.

For Rb--Cs and Rb--Rb pairs, the resulting error channels are \(\mathcal E_{\rm RbCs}^{\varphi}\) and \(\mathcal E_{\rm RbRb}^{\varphi}\), respectively, which incorporate both leakage and phase errors.
We quantify leakage by the probability that at least one atom is outside the computational subspace at the end of $\rm{CP}(\varphi)$,
\begin{equation}
p_{\mathrm{leak},q}^{\varphi}
=1-\frac{1}{4}
\sum_{k,l\in\mathcal B^{(2)}}
\langle l|
\mathcal E_{q}^{\varphi}
(|k\rangle\langle k|)
|l\rangle,
\label{eq:physical_error_rate_from_leakage}
\end{equation}
where \(\mathcal B^{(2)}=\{00,01,10,11\}\) is the two-qubit computational basis and $q\in{ \{\rm{RbCs},\rm{RbRb}\} }$.
In addition to leakage, we also have phase errors.
The phase error probabilities form a vector \(\boldsymbol p_{Z,q}^{\varphi}\equiv(p_{IZ},p_{ZI},p_{ZZ})\).

Under the global Rydberg pulse, idling atoms also undergo Rabi oscillations between \(\lvert1\rangle\) and \(\lvert r\rangle\) and are therefore subject to Rydberg-state decay. 
Solving the single-atom master equation gives the crosstalk error channel \(\mathcal E_s^{\varphi}\) for an idling atom of species \(s\in\{\mathrm{Rb},\mathrm{Cs}\}\).
The corresponding error probability is
\begin{equation}
\begin{aligned}
p_{\mathrm{cr},s}^{\varphi}
&=1-\frac{1}{2}\sum_{k,l=0}^{1}
\langle l|
\mathcal E_s^{\varphi}
(|k\rangle\langle k|)
|l\rangle.
\end{aligned}
\label{eq:crosstalk_depolarizing_conversion}
\end{equation}

In our circuit-level noise model for benchmarking fault tolerance, the effect of leakage is approximated by a two-qubit depolarizing channel \(\mathcal D_2\)~\cite{Suchara2015Leakage,Bluvstein2026}, so the full error channel of an $CP(\varphi)$ atom pair is therefore
\begin{equation}
\mathcal N_{q}^{\varphi}
=\mathcal D_2(p_{\mathrm{leak},q}^{\varphi})\circ\mathcal P_Z\!\left(\boldsymbol p_{Z,q}^{\varphi}\right).
\label{eq:native_rbcs_circuit_channel}
\end{equation}
Crosstalk is approximated by a single-qubit depolarizing channel \(\mathcal D_1(p_{\mathrm{cr},s}^{\varphi})\). 
Throughout our circuit-level simulations, we use \(p\equiv p_{\mathrm{leak},\mathrm{RbCs}}^{\pi}\) as the physical error rate.
We assign an error probability of \(p/10\) to microwave rotations and \(p\) to both reset and measurement~\cite{Sheng2018Microwave,Nikolov2023Microwave,highfidelity,Gidney2021faulttolerant}.
At each Rydberg Rabi frequency \(\Omega\), the resulting error channels together define our hardware-inferred circuit-level noise model, as summarized in Extended Data Table~\ref{tab:noise_model}.

\section{Surface Code Memory}
\label{app:memory_fit}

To benchmark the ZXXZ surface code compiled into \(\mathsf{GRAIL}\), we perform \(d\) rounds of syndrome extraction (SE), with logical state preparation and final readout in the \(\bar X\) and \(\bar Z\) bases.
To enable SE under global quantum control, we shuttle the measure qubits (Rb atoms) around the data qubits to extract the stabilizers (Fig.~\ref{fig:memory_performance}).
Exploiting the ZXXZ stabilizer geometry, we design a parallel shuttling scheme using crossed two-dimensional acousto-optic deflectors (2D-AODs)~\cite{Bluvstein2022CoherentTransport,logicals2026}. 
The resulting schedule requires eight parallel shuttling layers per SE round, independent of the code distance \(d\).

Logical state preparation and readout require addressed $H$ gates on selected Cs data qubits. 
For their compilation in $\mathsf{GRAIL}$, additional Rb addressors are introduced (Fig.~\ref{fig:memory_performance}\textbf{a}). 
The addressors are placed near Cs data qubits (within the Rydberg blockade radius) at specific positions \((x_j,y_j)\). 
For the $\bar X$- and $\bar Z$-basis tests, the positions take even and odd $(x_j+y_j)$, respectively.
The corresponding subsets are denoted as $\mathcal T_{\mathrm{Cs}}^{(\bar X)}$ and $\mathcal T_{\mathrm{Cs}}^{(\bar Z)}$, and the addressed \(H\) gates are $U_H(\mathcal T_{\mathrm{Cs}})$.

For physical realization, the SE time is determined by reset time \(T_{\mathcal R}\), readout time \(T_{\mathcal M}\), Rb one-site shuttling time $\tau$, microwave period \(T_{\mathrm{mw}}\),
\begin{equation}
t_{\mathrm{SE}}(\Omega/2\pi)
=T_{\mathcal R}+T_{\mathcal M}+8\tau+T_{\mathrm{mw}}+4.8T_\Omega.
\label{eq:memory_se_round_latency}
\end{equation}
With experimentally accessible parameters \(\Omega/2\pi=2.38\,\mathrm{MHz}\), \(T_{\mathcal R}=200\,\mu\mathrm{s}\), \(T_{\mathcal M}=50\,\mu\mathrm{s}\), $\tau=100\mu s$ and \(T_{\mathrm{mw}}=20\mu s\)~\cite{Zhang2024CavityDarkMode,Norcia2023,Zeng2026FastReadout}, the total SE duration is \(1.07\,\mathrm{ms}\).

Finally, we estimate the atom number cost required for the complete memory benchmark. 
For a distance-\(d\) surface code memory in logical basis \(B\in\{\bar X,\bar Z\}\), this comprises \(d^2\) Cs data qubits, \(d^2-1\) Rb measure qubits and \(N_{\mathrm{addr}}^{(B)}=|\mathcal T_{\mathrm{Cs}}^{(B)}|\) Rb addressors.
For \(\bar X\)- and \(\bar Z\)- basis tests, the required addressor counts are $((d^2 +1)/2)$ and $((d^2 -1)/2)$, respectively.
For \(d=17\), we require \(722\) and \(721\) atoms for \(\bar X\) and \(\bar Z\) benchmarks, respectively.

Supplementary Note~2 details the sampling, decoding and statistical analysis.

\section{Fold-transversal S gates with GRAIL}
\label{app:fold_transversal_gates}

The fold-transversal construction of the logical $\bar S$ gate involves addressed $H$, $S$, and $S^\dagger$ gates within a modified SE round~\cite{logicals2026}.
In $\mathsf{GRAIL}$, the implementation of these operations requires two global Rb resets, which would also erase the quantum information stored in the participating Rb syndrome ancillas.
To preserve this information, we temporarily transfer the states of these ancillas to additional Cs storage atoms and subsequently restore them after the addressed operations.
These transfers are implemented through quantum SWAP gates, rather than physical exchanges of atoms, and are denoted by $\mathrm{SWAP}_{\mathrm{in}}$ and $\mathrm{SWAP}_{\mathrm{out}}$, respectively.

Specifically, the following sequence is inserted halfway through an SE round:
\begin{equation}
\begin{aligned}
&\mathrm{SWAP}_{\mathrm{in}} \to \mathcal R_{\mathrm{Rb}} \\
&\to U_H \to U_S \to U_H \\
&\to \mathcal R_{\mathrm{Rb}}
 \to G_{\mathrm{Rb}}(H)
 \to \mathrm{SWAP}_{\mathrm{out}}.
\end{aligned}
\label{eq:logical_s_midpoint_window}
\end{equation}
Here, $U_H$ denotes the addressed $H$ layer required by the
ZXXZ encoding, while $U_S$ denotes the fold-transversal operation required by logical $\bar S$~\cite{logicals2026}.
Both $\mathcal R_{\mathrm{Rb}}$ operations are global resets and serve distinct purposes:
the first initializes fresh Rb addressors, while the second prepares the Rb measure qubits for the subsequent syndrome readout.

The logical Clifford benchmark circuits and statistical analysis are specified in Supplementary Note~3.

\section{Logical \texorpdfstring{\(\bar H_{XY}\)}{} Check in Magic State Cultivation}
\label{app:logical_check}

To implement the logical $\bar H_{XY}$ check, we first prepare a GHZ state among the Rb ancillas by applying two layers of global CZ gates, following the LOCC-based construction.
In the standard protocol, this is followed by a measurement and an outcome-dependent $X$ feedback to complete the construction.
However, in $\mathsf{GRAIL}$, the global Rb measurement would also project the Rb ancillas carrying the GHZ state, destroying the required entanglement.
To avoid this, we defer the measurements and associated feedback until after the controlled-\(\bar H_{XY}\) operation. 
The feedforward \(X\) corrections on the Rb ancillas are then propagated through the controlled gates, yielding corresponding corrections on the Cs data qubits.
The reordered sequence for GHZ preparation and the controlled-$\bar{H}_{XY}$ gate is
\begin{equation}
\begin{aligned}
&\mathcal R_{\mathrm{Rb}} \to G_{\mathrm{Rb}}(H) \to \mathrm{CZ}_{1} \to \mathrm{CZ}_{2} \\
&\to \mathrm{C}\text{-} \bar{H}_{XY} \to G_{\mathrm{Rb}}(H) \to \mathcal M_{\mathrm{Rb}} \to \text{Feedback}.
\end{aligned}
\label{eq:msc_logical_check_sequence}
\end{equation}
Here, the first line entangles the ancilla for the GHZ state preparation following the standard LOCC scheme.
Next, we immediately apply the controlled-$\bar{H}_{XY}$ gate before the final global measurement and feedback. 
Another key difference is that we use a GHZ ancilla of size $d^2$ instead of $d$ as in the original MSC~\cite{sahay2026foldmsc}. 
This allows for parallel implementation of physical gates in the controlled-$\bar{H}_{XY}$ under global control.

\endgroup

\small
\noindent\textbf{Data availability}\\
The data that support the findings of this article are openly available.

\medskip
\noindent\textbf{Code availability}\\
All code used to generate the numerical results and figures is available upon reasonable request.

\medskip
\noindent\textbf{Acknowledgements}\\
We thank Peter Zoller, Hannes Bernien and Dongling Deng for helpful discussions. 
This work is supported by the National Program on Key Basic Research Project of China (Grant No. 2021YFA1400900), the Innovation Program for Quantum Science and Technology of China (Grant No. 2024ZD0300100), the Shanghai Municipal Science and Technology Major Project (Grant Nos. 26LZ0500100, 25TQ003, 24LZ1400900 and 24DP2600100), the Shanghai Municipal Science and Technology Co-research Program (Grant No. 25LZ2601000), and the Scientific Research Innovation Capability Support Project for Young Faculty (Grant No. ZYGXQNJSKYCXNLZCXM-I14).

\medskip
\noindent\textbf{Author contributions}\\
X.L. and L.F. conceived and supervised the project. Z.Z. and S.C. developed the quantum circuit realization and carried out the numerical simulations.  All authors contributed to the $\mathsf{GRAIL}$ architecture design, the physical realization, and the writing of the manuscript. \\

\medskip
\noindent\textbf{Competing interests}\\
X.Y., Y.M., Y.L., L.F., and X.L. are shareholders of
Shanghai Buchou Quantum Technology Co., Ltd.
Shanghai Buchou Quantum Technology Co., Ltd and Fudan University
have filed the following Chinese patent applications,
naming X.L., Z.Z. and X.Y. as inventors:
(i) A dual-species quantum computing system and a method for its control
(application No. 202611675318.9);
(ii) A method for synthesizing physical gates in a dual-species system
(application No. 202611675346.0);
(iii) A method for storing quantum information using error-correcting
codes and extracting syndromes
(application No. 202611675369.1);
(iv) A method for compiling and executing fault-tolerant logical
Clifford gates
(application No. 202611675379.5); and
(v) A method for processing non-Clifford resource states
(application No. 202611675388.4).
The applications were filed on 27 October 2026 and accepted
for processing by the China National Intellectual Property Administration.
The remaining authors declare no competing interests.

\medskip
\noindent\textbf{Additional information}\\
Supplementary information accompanies this paper. 
Correspondence and requests for materials should be addressed to the corresponding authors.

\clearpage
\begingroup
\setcounter{figure}{0}
\setcounter{table}{0}
\renewcommand{\figurename}{Extended Data Fig.}
\renewcommand{\tablename}{Extended Data Table}
\renewcommand{\thefigure}{\arabic{figure}}
\renewcommand{\thetable}{\arabic{table}}
\renewcommand*{\theHfigure}{extendeddata.figure.\arabic{figure}}
\renewcommand*{\theHtable}{extendeddata.table.\arabic{table}}

\onecolumngrid

\begin{figure}[p]
\centering
\includegraphics[width=0.48\textwidth]{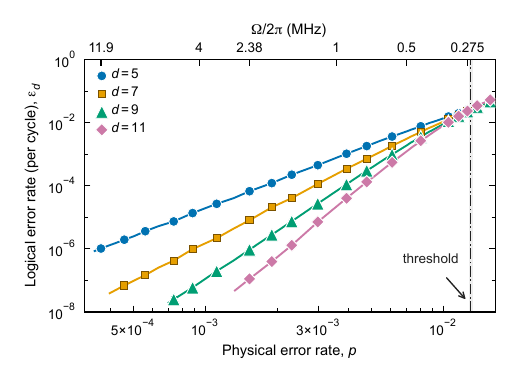}
\caption{\textbf{Memory performance without crosstalk (NCR).}
The Cs and Rb crosstalk channels are removed from the full hardware-inferred circuit-level noise model.
The logical error rates per cycle \(\varepsilon_d\) are averaged over the \(\bar X\) and \(\bar Z\) memory tests after \(d\) SE rounds.
Error bars show 95\% confidence intervals. 
The threshold is \(p_{\mathrm{th}}=1.30\times10^{-2}\).
The error suppression factor is \(\Lambda=8.47\pm0.03\), where the uncertainty is one fit standard error.
The threshold and error suppression factor appear to be close to those of the full noise model. 
The number of independent Monte Carlo shots per data point ranges from \(10^4\) to \(10^9\) across the two logical bases combined.}
\label{fig:memory_performance_no_cr}
\end{figure}

\clearpage

\clearpage

\begin{table}[t]
\centering
\caption{\textbf{The hardware-inferred circuit-level noise model.}
The channel definitions are given in \methodsref{app:heterogeneous_noise}.
}
\label{tab:noise_model}
\small
\begin{tblr}{
width=\linewidth,
colspec={X[1.05,c] X[1.20,c] X[1.50,c] X[1.35,c]},
cells={halign=c,valign=m},
cell{2}{1}={r=2}{},
row{1}={font=\bfseries,halign=c},
colsep=5pt,
rowsep=4pt,
hline{1,7}={0.08em,solid},
hline{2,4,5,6}={0.04em,solid},
}
Native operation
& Affected atoms
& Error channel
& Error probabilities \\

Global Rydberg pulse \(\mathrm{Ryd}_s(\varphi)\)
& Participating atoms
& \(\begin{gathered}
\mathcal D_2\!\left(p_{\mathrm{leak},q}^{\varphi}\right)
\circ\mathcal P_Z\!\left(\boldsymbol p_{Z,q}^{\varphi}\right)
\end{gathered}\)
& \(\begin{gathered}
p_{\mathrm{leak},q}^{\varphi}(\Omega)\\
\boldsymbol p_{Z,q}^{\varphi}(\Omega)
\end{gathered}\) \\

& Idling atoms
& \(\mathcal D_1\!\left(p_{\mathrm{cr},s}^{\varphi}\right)\)
& \(p_{\mathrm{cr},s}^{\varphi}(\Omega)\) \\

Global microwave rotation \(G_s(U)\)
& Every atom in \(\mathcal A_s\)
& {\(\mathcal D_1(p_{\mathrm{mw}})\)\\ }
& \(p_{\mathrm{mw}}=p/10\) \\

Global reset \(\mathcal R_s\)
& Every atom in \(\mathcal A_s\)
& Bit flip
& \(p_{\mathrm{reset}}=p\) \\

Global measurement \(\mathcal M_s\)
& Every atom in \(\mathcal A_s\)
& Bit flip
& \(p_{\mathrm{meas}}=p\) \\
\end{tblr}
\end{table}

\clearpage
\endgroup

\end{document}


\maketitle

\begingroup
\renewcommand{\abstractname}{Overview}
\begin{abstract}
This Supplementary Information provides implementation details and numerical methods for quantum circuits compiled into \(\mathsf{GRAIL}\).
Supplementary Note~\ref{sec:si_physical_gates} specifies the Rydberg pulse parameters and phase corrections, estimates physical gate durations, verifies gate decompositions and describes the calculation of gate and crosstalk fidelities.
Supplementary Note~\ref{sec:si_memory_benchmark} presents the sampling, decoding and statistical analysis of surface-code memory benchmarks, together with threshold estimation, below-threshold scaling and distance extrapolation.
Supplementary Note~\ref{sec:si_clifford_benchmarks} describes the logical \(\bar H\), \(\bar S^2\) and \(\overline{\mathrm{CZ}}\) benchmarks, including their circuits, atom counts, decoding methods, error-rate estimates, thresholds and distance extrapolations.
Supplementary Note~\ref{sec:si_magic_state_cultivation} details the \(f=5\) \(T\)-state cultivation sequence, logical checks and postponed feedback, followed by the decoding, postselection and statistical analysis used to quantify stage survival, logical error rates and the expected number of attempts per accepted output.
\end{abstract}
\endgroup

\clearpage
\tableofcontents
\clearpage

\section{Physical gates under global control}
\label{sec:si_physical_gates}

Here, we specify the Rydberg pulse used in \(\mathsf{GRAIL}\) and estimate the physical-gate durations.
We then verify the decompositions of addressed single-qubit gates, Rb-mediated two-qubit gates, Rb--Cs SWAP gates and parallelized CCZ gates, including their action on addressors and idling atoms.
Finally, we present the expressions used to evaluate gate and crosstalk fidelities.

\subsection{Rydberg pulse parameters and phase correction}
\label{sec:si_atomic_dynamics}
\label{sec:single_atom_simulation}

The global Rydberg drives on Rb and Cs have the same constant Rabi frequency \(\Omega\) and phase profile~\cite{highfidelity},
\begin{equation}
\phi(t)=A\cos(\omega_0t-\phi_0)+\delta_0t,
\label{eq:rydberg_phase_profile}
\end{equation}
where \(A\) and \(\omega_0\) are the amplitude and angular frequency of the phase modulation, \(\phi_0\) is its phase offset, and \(\delta_0\) is the detuning. 
We optimize the pulse parameters using Powell's method~\cite{Powell1964}.
The optimized values are listed in Supplementary Table~\ref{tab:rbcs_pulse_parameters}.
The resulting pulse implements 
\begin{equation}
U_{\mathrm{opt}}(\varphi)
=[P(\theta)\otimes P(\theta)]\mathrm{CP}(\varphi),
\end{equation}
where \(P(\theta)=\operatorname{diag}(1,e^{i\theta})\) and \(\theta\) is a fixed single-qubit phase. 
An idling atom undergoes \(P(\theta)\). 
We compensate these unwanted single-qubit phases using virtual-\(Z\) rotations~\cite{Nikolov2023Microwave}.
In the main text, these phase corrections are included by default in the definition of the global Rydberg pulse \(\mathrm{Ryd}_s(\varphi)\).

\begin{table}[htbp]
\centering
\caption{\textbf{Optimized Rydberg pulse parameters and single-qubit phases.}
The pulses implement \(\mathrm{CP}(\varphi)\) gates.
\(\kappa=\Omega T_{\mathrm{gate}}/(2\pi)\) is the normalized pulse duration.
The optimized pulse parameters and corresponding single-qubit phases are rounded for display.
The unrounded values used in the calculations are provided in the reproducibility package.
}
\label{tab:rbcs_pulse_parameters}
\begin{tabular}{lccc}
\toprule
Parameter
& \(\mathrm{CZ}\equiv\mathrm{CP}(\pi)\)
& \(\mathrm{CP}(\pi/2)\)
& \(\mathrm{CP}(\pi/4)\) \\
\midrule
\(\kappa\)
& \(1.2\) & \(1.1\) & \(1.0\) \\
\(A\) (rad)
& \(0.58795\) & \(0.85786\) & \(0.53313\) \\
\(\omega_0/\Omega\)
& \(1.2743\) & \(1.3487\) & \(1.7576\) \\
\(\delta_0/\Omega\)
& \(-0.058159\) & \(0.47299\) & \(0.65176\) \\
\(\phi_0\) (rad)
& \(-3.0490\) & \(-0.051501\) & \(-2.3323\) \\
\(\theta\) (rad)
& \(-3.9318\) & \(-0.75514\) & \(-1.4928\) \\
\bottomrule
\end{tabular}
\end{table}

\subsection{Physical-gate durations}
\label{sec:si_physical_gate_timing}

We estimate the durations of physical gates on our array using typical experimental parameters~\cite{Bluvstein2022CoherentTransport,Zhou2025TransversalResources}.
We use the Rydberg Rabi period \(T_\Omega=(\Omega/2\pi)^{-1}\), a microwave \(2\pi\) rotation time \(T_{\mathrm{mw}}=20\,\mu\mathrm{s}\), and a time \(\tau=100\,\mu\mathrm{s}\) for moving an Rb atom by one lattice site using an acousto-optic deflector (AOD).
Following the gate sequences in Fig.~1\textbf{b,c} in the main text, the resulting duration estimates are given in Supplementary Table~\ref{tab:si_physical_gate_timing}.

\begin{table}[H]
\centering
\small
\caption{\textbf{Estimated physical-gate durations.}
The gate durations are dominated by atom shuttling and therefore depend weakly on the Rydberg Rabi frequency.
The three numerical estimates are evaluated at \(\Omega/2\pi=2.38\,\mathrm{MHz}\).}
\label{tab:si_physical_gate_timing}
\begin{tabular}{@{}lccc@{}}
\toprule
 & \(H\) & \(T\) & \(\mathrm{CZ}\) \\
\midrule
Duration
 & \(1.2T_\Omega+T_{\mathrm{mw}}/4+2\tau\)
 & \(T_\Omega+2\tau\)
 & \(3.6T_\Omega+T_{\mathrm{mw}}+6\tau\) \\
Estimate (\(\mu\mathrm{s}\)) & \(205.5\) & \(200.4\) & \(621.5\) \\
\bottomrule
\end{tabular}
\end{table}

\subsection{Physical gate decompositions}
\label{sec:physical_gate_decompositions}

Here, we verify the physical gate decompositions.

\paragraph{Addressed single-qubit gate.}
For a single-qubit gate on Cs qubit \(j\) (Fig.~1\textbf{b} of the main text), the first global Rb \(X\) gate maps the associated Rb addressor \(i\) from \(\lvert0\rangle\) to \(\lvert1\rangle\).
The subsequent controlled-phase gate therefore acts as
\begin{equation}
\mathrm{CP}_{i,j}(\varphi)
\bigl(\lvert1\rangle_i\otimes\lvert\psi\rangle_j\bigr)
=\lvert1\rangle_i\otimes P_j(\varphi)\lvert\psi\rangle_j .
\label{eq:addressor_cp_action}
\end{equation}
Combined with the surrounding global Cs rotations, this phase gate implements the single-qubit unitary \(U=R_\mu(-\alpha)P(\varphi)R_\mu(\alpha)\) defined in the main text, up to an overall phase.
The final global Rb \(X\) gate returns the addressor to \(\lvert0\rangle_i\).
An idling Cs or Rb atom undergoes \(R_\mu(-\alpha)IR_\mu(\alpha)=I\) or \(XIX=I\), respectively, preserving arbitrary input states.

\paragraph{Rb-mediated two-qubit gate.}
For a two-qubit gate on a Cs pair \((j_1,j_2)\), the shared Rb addressor \(i\) starts in \(\lvert0\rangle\) and interacts successively with \(j_1\), \(j_2\) and \(j_1\).
We use the identity \(H_i\mathrm{CZ}_{i,j_1}H_i=\mathrm{CX}_{j_1\to i}\), where the arrow points from control to target.
Let \(x,y\in\{0,1\}\) be the bit values of Cs qubits \(j_1\) and \(j_2\), respectively.
With the qubits ordered as \((i,j_1,j_2)\), the sequence acts as
\begin{equation}
\begin{aligned}
\lvert0,x,y\rangle
&\xrightarrow{\mathrm{CX}_{j_1\to i}}\lvert x,x,y\rangle\\
&\xrightarrow{\mathrm{CP}_{i,j_2}(\varphi)}e^{i\varphi xy}\lvert x,x,y\rangle\\
&\xrightarrow{\mathrm{CX}_{j_1\to i}}e^{i\varphi xy}\lvert0,x,y\rangle .
\end{aligned}
\label{eq:physical_cp_basis_action}
\end{equation}
The sequence therefore implements \(\mathrm{CP}_{j_1,j_2}(\varphi)\) on any input state of the Cs pair and returns the addressor to \(\lvert0\rangle_i\), disentangled from the data qubits.
An idling Rb atom undergoes \(H^4=I\), and an idling Cs atom experiences the identity.

\paragraph{SWAP gate between Rb and Cs.}
For an Rb atom \(a\) and a Cs atom \(b\), we implement \(\mathrm{SWAP}_{a,b}\) using three native gates \(\mathrm{CZ}_{a,b}\).
The sequence is
\begin{equation}
\begin{aligned}
&G_{\mathrm{Cs}}(H)\to\mathrm{CZ}_{a,b}\to G_{\mathrm{Cs}}(H)\\
&\to G_{\mathrm{Rb}}(H)\to\mathrm{CZ}_{a,b}\to G_{\mathrm{Rb}}(H)\\
&\to G_{\mathrm{Cs}}(H)\to\mathrm{CZ}_{a,b}\to G_{\mathrm{Cs}}(H).
\end{aligned}
\label{eq:si_rbcs_swap_sequence}
\end{equation}
For a computational-basis input ordered as \(a,b\), the sequence acts as
\begin{equation}
\begin{aligned}
\lvert x,y\rangle
&\xrightarrow{\mathrm{CX}_{a\to b}}\lvert x,x\oplus y\rangle\\
&\xrightarrow{\mathrm{CX}_{b\to a}}\lvert y,x\oplus y\rangle\\
&\xrightarrow{\mathrm{CX}_{a\to b}}\lvert y,x\rangle .
\end{aligned}
\label{eq:si_rbcs_swap_basis_action}
\end{equation}
The sequence therefore exchanges the states of the two atoms for any joint input state.
An idling Rb atom undergoes \(H^2=I\), and an idling Cs atom undergoes \(H^4=I\).

\paragraph{Parallelized CCZ gate scheme.}
We present a parallelized implementation of CCZ gates on Rb--Cs--Cs triples. 
Each gate acts on an Rb control \(c_j\) and two Cs targets \(a_j\) and \(b_j\), with each selected qubit participating in one CCZ gate.
To mediate the interaction between the two Cs targets, we assign a separate Rb addressor \(m_j\) to each gate.
The native CP layers use the following Rb--Cs pair sets:
\begin{equation}
\begin{aligned}
\mathcal Q_{ca}&=\{(c_j,a_j)\}_j,
&\mathcal Q_{cb}&=\{(c_j,b_j)\}_j,\\
\mathcal Q_{ma}&=\{(m_j,a_j)\}_j,
&\mathcal Q_{mb}&=\{(m_j,b_j)\}_j,
\end{aligned}
\label{eq:msc_ccz_pair_sets}
\end{equation}
where \(j\) runs over the selected CCZ gates.
The parallelized sequence consists of nine such layers, 
\begin{equation}
\begin{aligned}
&\quad \ \ \mathrm{CP}(\mathcal Q_{ca};\pi/2)\\
&\to\mathrm{CP}(\mathcal Q_{cb};\pi/2)\\
&\to G_{\mathrm{Cs}}(H)\to\mathrm{CP}(\mathcal Q_{mb};\pi)\to G_{\mathrm{Cs}}(H)\\
&\to G_{\mathrm{Rb}}(H)\to\mathrm{CP}(\mathcal Q_{ma};\pi)\to G_{\mathrm{Rb}}(H)\\
&\to G_{\mathrm{Cs}}(H)\to\mathrm{CP}(\mathcal Q_{mb};\pi)\to G_{\mathrm{Cs}}(H)\\
&\to\mathrm{CP}(\mathcal Q_{cb};-\pi/2)\\
&\to G_{\mathrm{Cs}}(H)\to\mathrm{CP}(\mathcal Q_{mb};\pi)\to G_{\mathrm{Cs}}(H)\\
&\to G_{\mathrm{Rb}}(H)\to\mathrm{CP}(\mathcal Q_{ma};\pi)\to G_{\mathrm{Rb}}(H)\\
&\to G_{\mathrm{Cs}}(H)\to\mathrm{CP}(\mathcal Q_{mb};\pi)\to G_{\mathrm{Cs}}(H).
\end{aligned}
\label{eq:msc_ccz_native_sequence}
\end{equation}
Here, inverse-pulse symmetry relates \(\mathrm{CP}(-\pi/2)\) to \(\mathrm{CP}(\pi/2)\), reversing the single-qubit phase corrections while preserving the same error error model.
For gate \(j\), write the computational-basis state of \(c_j,a_j,b_j,m_j\), in that order, as \(\lvert z,x,y,w\rangle\).
The first two rows contribute the phase \(e^{i\pi z(x+y)/2}\).
Rows 3--5 implement \(\mathrm{CX}_{m_j\to b_j}\), \(\mathrm{CX}_{a_j\to m_j}\) and \(\mathrm{CX}_{m_j\to b_j}\), so the bits on \(b_j\) and \(m_j\) evolve as
\begin{equation}
(y,w)
\longmapsto (y\oplus w,w)
\longmapsto (y\oplus w,w\oplus x)
\longmapsto (x\oplus y,w\oplus x).
\end{equation}
With \(b_j\) now holding \(x\oplus y\), row 6 contributes the phase \(e^{-i\pi z(x\oplus y)/2}\).
Rows 7--9 restore the original bits \(y\) and \(w\) on \(b_j\) and \(m_j\).
Since \(x+y-(x\oplus y)=2xy\), the full sequence gives
\begin{equation}
\begin{aligned}
\lvert z,x,y,w\rangle
&\longmapsto e^{i\pi z[x+y-(x\oplus y)]/2}\lvert z,x,y,w\rangle\\
&=(-1)^{zxy}\lvert z,x,y,w\rangle.
\end{aligned}
\label{eq:si_ccz_basis_action}
\end{equation}
Thus the sequence implements \(\prod_j\mathrm{CCZ}_{c_j,a_j,b_j}\) on the selected triples and acts as the identity on every addressor, so an addressor initialized in \(\lvert0\rangle\) returns to \(\lvert0\rangle\).
The net operation on every idling atom is also the identity, since the global microwave rotations give \(H^4=I\) on Rb and \(H^8=I\) on Cs.

\subsection{Gate and crosstalk fidelity}
\label{sec:si_average_fidelity}

Here, we present the expressions used to calculate the gate and crosstalk fidelities reported in Section~2 of the main text.

Let \(\mathcal E\) denote the noisy channel, \(U\) the ideal unitary, and \(\Pi_q\) the projector onto the \(d_q\)-dimensional computational subspace \(\mathcal H_q\).
The average fidelity is defined as~\cite{fidelity1999,Nielsen2002}
\begin{equation}
\mathcal F
=\mathbb E_\psi\!\left[
\langle\psi|U^\dagger\mathcal E(|\psi\rangle\langle\psi|)U|\psi\rangle
\right],
\label{eq:haar_fidelity_definition}
\end{equation}
where \(\mathbb E_\psi\) denotes the Haar average over pure states in \(\mathcal H_q\).
For an orthonormal basis \(\{|i\rangle\}\) of \(\mathcal H_q\), we define the average probability of remaining in the computational subspace as
\begin{equation}
p_q=\frac{1}{d_q}\sum_i\operatorname{Tr}\!\left[\Pi_q\mathcal E(|i\rangle\langle i|)\right],
\label{eq:isometry_survival}
\end{equation}
and the entanglement fidelity as
\begin{equation}
\mathcal F_{\mathrm e}
=\frac{1}{d_q^2}\sum_{i,j}\langle i|U^\dagger\mathcal E(|i\rangle\langle j|)U|j\rangle.
\label{eq:isometry_entanglement_fidelity}
\end{equation}
The average fidelity is then given by~\cite{WoodGambetta2018,fidelity2021}
\begin{equation}
\mathcal F=\frac{d_q\mathcal F_{\mathrm e}+p_q}{d_q+1},
\label{eq:isometry_haar_fidelity}
\end{equation}
with \(d_q=2\) for the \(H\) and \(T\) gates and for single-atom crosstalk, and \(d_q=4\) for the \(\mathrm{CZ}\) gate.

\section{Surface-code memory benchmark}
\label{sec:si_memory_benchmark}

Here, we present numerical benchmarks of surface-code memory circuits compiled into native \(\mathsf{GRAIL}\) instruction sequences.
We first describe the sampling and decoding procedures and the estimation of statistical uncertainties for benchmarks under the hardware-inferred circuit-level noise model with crosstalk (CR) and the corresponding model without crosstalk (NCR).
We then estimate the error-correction thresholds, characterize the dependence of the logical error rate on code distance and extrapolate the distance required to achieve a logical error rate below \(10^{-9}\) per cycle.

\subsection{Sampling, decoding and statistical uncertainty}
\label{sec:si_memory_sampling}

For the CR and NCR memory benchmarks shown in Fig.~3 of the main text and Extended Data Fig.~1, respectively, we sample the circuits at the same 46 Rydberg Rabi frequencies, with \(\Omega/2\pi\) ranging from \(0.2\) to \(12\,\mathrm{MHz}\).
We use the minimum-weight perfect matching decoder implemented in \texttt{PyMatching}~\cite{pymatching}.

For each logical basis \(B\in\{\bar X,\bar Z\}\), the benchmark failure probability and effective logical error per cycle are
\begin{equation}
\begin{aligned}
p_{\mathrm L,d}^{(B)}(r)
&=\frac{N_{\mathrm{fail},B}}{N_{\mathrm{shots},B}},\\
\varepsilon_d^{(B)}
&=\frac{1-\left[1-2p_{\mathrm L,d}^{(B)}(r)\right]^{1/r}}{2},
\qquad r=d.
\end{aligned}
\label{eq:memory_per_cycle_definition}
\end{equation}
We report equally weighted \(\varepsilon_d=(\varepsilon_d^{(\bar X)}+\varepsilon_d^{(\bar Z)})/2\).
We obtain the error bars on \(\varepsilon_d\) from the two-sided \(95\%\) confidence intervals for the \(\bar X\)- and \(\bar Z\)-basis failure probabilities.

\subsection{Threshold estimation and distance extrapolation}
\label{sec:si_memory_fitting}

The error-correction threshold characterizes the physical noise scale below which the logical error rate decreases exponentially with increasing code distance.
In our circuit-level simulations, we take \(p\equiv p_{\mathrm{leak},\mathrm{RbCs}}^\pi\) as the physical error rate.
From the memory performance, we estimate the threshold as the crossing point of the logical error rate curves.
This gives a CR threshold of \(1.43\%\), with a \(95\%\) confidence interval of \([1.40,1.47]\%\) (Fig.~3\textbf{c} of the main text), and an NCR threshold of \(1.30\%\), with a \(95\%\) confidence interval of \([1.28,1.33]\%\) (Extended Data Fig.~1).
Lindblad simulations of the native Rb--Cs \(\mathrm{CZ}\) gate give the leakage probability \(p(\Omega)\) as a function of the Rydberg Rabi frequency \(\Omega\).
Interpolating these numerical results gives the corresponding threshold frequencies \(\Omega_{\mathrm{th}}/2\pi=0.242\,\mathrm{MHz}\) for CR and \(0.267\,\mathrm{MHz}\) for NCR, with \(95\%\) confidence intervals of \([0.236,0.248]\,\mathrm{MHz}\) and \([0.260,0.271]\,\mathrm{MHz}\), respectively.

To characterize the expected exponential decrease of the logical error rate with increasing code distance, we fit the below-threshold CR memory data (\(p<0.01\)) to the form~\cite{Fowler2012SurfaceCodes,ORourke2024ComparePair}
\begin{equation}
\varepsilon_d=\alpha\left(p/\beta\right)^{\gamma d-\delta}.
\end{equation}
We obtain the fit parameters \(\alpha=0.0413\) with an interval of \([0.0411,0.0414]\),
\(\beta=0.01263\) with an interval of \([0.01262,0.01264]\),
\(\gamma=0.4819\) with an interval of \([0.4812,0.4826]\), and
\(\delta=-0.356\) with an interval of \([-0.360,-0.352]\).
Within this fitted model, \(\gamma>0\) and \(p<\beta\) imply that \(\varepsilon_d\) decreases exponentially with code distance.

We next estimate the code distance required for quantum memory to reach a logical error rate per cycle below \(10^{-9}\), motivated by the requirements of quantum algorithms~\cite{Viszlai2026QSIEVE,Cain2026ShorAtomicQubits,Gidney2017Factoring,Haner2020EllipticCurve,Babbush2018ElectronicSpectra,Gidney2025RSA2048,Zhou2025TransversalResources}.
Using the CR data at \(\Omega/2\pi=2.38\,\mathrm{MHz}\) for \(d=5,7,9,11\), we estimate the error suppression factor by fitting
\begin{equation}
\ln\varepsilon_d=C-\zeta d,
\qquad
\Lambda=e^{2\zeta}
=\frac{\varepsilon_d}{\varepsilon_{d+2}}.
\label{eq:memory_distance_scaling}
\end{equation}
The fit gives \(\Lambda_{\mathrm{CR}}=8.76\pm0.04\), where the uncertainty is one standard error from the fit.
Extrapolating this fit to larger code distances, we find that \(d=17\) is sufficient to achieve a mean logical error rate per cycle below \(10^{-9}\) at \(\Omega/2\pi=2.38\,\mathrm{MHz}\).
Applying the same fitting procedure to the NCR data gives \(\Lambda_{\mathrm{NCR}}=8.47\pm0.03\), with the uncertainty again representing one standard error from the fit.

\section{Logical Clifford benchmarks}
\label{sec:si_clifford_benchmarks}

We numerically benchmark logical Clifford gates using circuits compiled into native \(\mathsf{GRAIL}\) instruction sequences.
We first describe the benchmark procedures, atom counts, decoding methods and statistical analysis underlying the results reported in Fig.~4 of the main text.
We then estimate the error-correction thresholds and use fits to data sufficiently below threshold to extrapolate the code distances required to achieve logical error rates below \(10^{-9}\).

\subsection{Sampling, decoding and statistical uncertainty}
\label{sec:si_clifford_sampling}

Here, we describe the benchmark procedures for the three logical
Clifford gates, as summarized in
Supplementary Table~\ref{tab:logical_clifford_benchmarks}.

\begin{table}[htbp]
\centering
\small
\setlength{\tabcolsep}{4pt}
\renewcommand{\arraystretch}{1.15}
\caption{\textbf{Logical Clifford benchmark procedures and atom counts.}
An arrow \(\bar P\mapsto \bar Q\) denotes preparation in an eigenstate of the logical observable \(\bar P\), application of the gate, and readout of \(\bar Q\), with the ideal output eigenvalue equal to the input eigenvalue.
For pairs of observables, this convention applies componentwise.
Each mapping listed for \(\bar H\) and \(\overline{\mathrm{CZ}}\) is sampled separately.
For \(\overline{\mathrm{CZ}}\), the subscripts \(C\) and \(T\) label the two code patches, and a shot is counted as a single failure if either output observable has an incorrect eigenvalue.
The atom counts include Cs data qubits, Rb measure qubits and Rb addressors for one code patch (\(\bar H\) and \(\bar S\)) or two code patches (\(\overline{\mathrm{CZ}}\)).
For \(\bar S\), the Cs count additionally includes storage atoms that hold the syndrome information transferred from Rb syndrome ancillas during the fold-transversal operation \(U_S\).
}
\label{tab:logical_clifford_benchmarks}
\begin{tabular}{@{}
>{\raggedright\arraybackslash}p{0.09\textwidth}
>{\raggedright\arraybackslash}p{0.55\textwidth}
>{\centering\arraybackslash}p{0.13\textwidth}
>{\centering\arraybackslash}p{0.13\textwidth}
@{}}
\toprule
Gate
& Benchmark procedure
& Cs atoms
& Rb atoms \\
\midrule
\(\bar H\)
& \(\bar X\mapsto\bar Z\),  \(\bar Z\mapsto\bar X\)
& \(d^2\)
& \(2d^2-1\) \\
\addlinespace
\(\bar S\)
& Apply two \(\bar S\) gates to a \(\bar X\) eigenstate and measure \(\bar X\), whose eigenvalue ideally flips (\(\bar S^2=\bar Z\)).
& \(2d^2-d\)
& \(2d^2-1\) \\
\addlinespace
\(\overline{\mathrm{CZ}}\)
& \((\bar X_C,\bar Z_T)\mapsto
     (\bar X_C\bar Z_T,\bar Z_T)\),
  \((\bar Z_C,\bar X_T)\mapsto
     (\bar Z_C,\bar Z_C\bar X_T)\)
& \(2d^2\)
& \(4d^2-2\) \\
\bottomrule
\end{tabular}
\end{table}

For \(\bar H\), we use \texttt{PyMatching} to decode~\cite{pymatching}.
For \(\bar S^2\) and \(\overline{\mathrm{CZ}}\), we use belief propagation with ordered statistics decoding (BP+OSD) via \texttt{ldpc.BpOsdDecoder}~\cite{Roffe_LDPC_Python_tools_2022,roffe_decoding_2020}.

The logical error rate \(p_{\mathrm L,d}\) is the failure probability of the complete benchmark.
At a given code distance and Rydberg Rabi frequency, let \(k_b\) and \(N_b\) denote the number of failed shots and the total number of shots, respectively, for benchmark setting \(b\).
The reported logical error rates are
\begin{equation}
\begin{aligned}
p_{\mathrm L,d}^{(\bar H)}
&=\frac{1}{2}\left(k_X/N_X+k_Z/N_Z\right),\\
p_{\mathrm L,d}^{(\bar S^2)}
&=k_{S^2}/N_{S^2},\\
p_{\mathrm L,d}^{(\overline{\mathrm{CZ}})}
&=\frac{1}{2}\left(k_{XZ}/N_{XZ}+k_{ZX}/N_{ZX}\right),
\end{aligned}
\label{eq:si_clifford_error_rates}
\end{equation}
where \(X\) and \(Z\) label the two input bases for the \(\bar H\) benchmark, \(S^2\) labels the benchmark with two \(\bar S\) operations, and \(XZ\) and \(ZX\) label the two joint input bases for the \(\overline{\mathrm{CZ}}\) benchmark listed in Supplementary Table~\ref{tab:logical_clifford_benchmarks}.
We estimate the statistical uncertainty in the logical error rates using two-sided \(95\%\) confidence intervals.

\subsection{Threshold estimation and distance extrapolation}
\label{sec:si_clifford_fitting}

Here, we evaluate the fault-tolerant performance of the logical Clifford gates using the \(\bar H\), \(\bar S^2\) and \(\overline{\mathrm{CZ}}\) benchmarks shown in Fig.~4 of the main text.
From the performance of logical Clifford gates, we estimate the threshold as the crossing point of the logical error rate curves.
This gives a threshold of \(0.9749\%\), with a \(95\%\) confidence interval of \([0.9745,0.9754]\%\), for \(\bar H\);
\(0.523\%\), with a \(95\%\) confidence interval of \([0.521,0.526]\%\), for \(\bar S^2\);
and \(0.775\%\), with a \(95\%\) confidence interval of \([0.774,0.777]\%\), for \(\overline{\mathrm{CZ}}\).
We convert these thresholds to Rydberg Rabi frequencies by interpolating the leakage probabilities \(p(\Omega)\) obtained from Lindblad simulations of the native Rb--Cs \(\mathrm{CZ}\) gate.
The corresponding threshold frequencies \(\Omega_{\mathrm{th}}/2\pi\) are
\(0.3574\,\mathrm{MHz}\), with a \(95\%\) confidence interval of \([0.3572,0.3575]\,\mathrm{MHz}\);
\(0.673\,\mathrm{MHz}\), with a \(95\%\) confidence interval of \([0.669,0.676]\,\mathrm{MHz}\);
and \(0.451\,\mathrm{MHz}\), with a \(95\%\) confidence interval of \([0.450,0.452]\,\mathrm{MHz}\), respectively.

To characterize the below-threshold behavior, we fit the data at \(d=5,7,9\) using the same subthreshold scaling form as for quantum memory,
\begin{equation}
p_{\mathrm L,d}=\alpha\left(p/\beta\right)^{\gamma d-\delta}.
\label{eq:si_clifford_distance_extrapolation}
\end{equation}
For each benchmark, we restrict the fit to a range of \(p\) sufficiently below the estimated threshold to avoid near-threshold effects (\(p<0.007\), \(p<0.003\) and \(p<0.006\) for \(\bar H\), \(\bar S^2\) and \(\overline{\mathrm{CZ}}\), respectively).
The fitted parameters are summarized in Supplementary Table~\ref{tab:clifford_subthreshold_parameters}.

\begin{table}[htbp]
\centering
\small
\caption{\textbf{Fitted parameters of the subthreshold scaling model for the logical Clifford benchmarks.}
Each table cell gives the point estimate, with its two-sided \(95\%\) confidence interval in brackets below.
The \(\bar S\) column refers to the complete \(\bar S^2\) benchmark.}
\label{tab:clifford_subthreshold_parameters}
\begin{tabular}{lccc}
\toprule
Parameter & \(\bar H\) & \(\bar S\) & \(\overline{\mathrm{CZ}}\) \\
\midrule
\(\alpha\)
& \shortstack{\(0.1147\) \\ \([0.1141,0.1154]\)}
& \shortstack{\(0.1954\) \\ \([0.1882,0.2032]\)}
& \shortstack{\(0.2295\) \\ \([0.2233,0.2360]\)} \\
\addlinespace
\(\beta\)
& \shortstack{\(0.00903\) \\ \([0.00901,0.00904]\)}
& \shortstack{\(0.00454\) \\ \([0.00449,0.00460]\)}
& \shortstack{\(0.00715\) \\ \([0.00708,0.00722]\)} \\
\addlinespace
\(\gamma\)
& \shortstack{\(0.499\) \\ \([0.497,0.500]\)}
& \shortstack{\(0.466\) \\ \([0.456,0.477]\)}
& \shortstack{\(0.477\) \\ \([0.470,0.485]\)} \\
\addlinespace
\(\delta\)
& \shortstack{\(-0.298\) \\ \([-0.306,-0.289]\)}
& \shortstack{\(-0.404\) \\ \([-0.462,-0.345]\)}
& \shortstack{\(-0.448\) \\ \([-0.489,-0.408]\)} \\
\bottomrule
\end{tabular}
\end{table}

For all three benchmarks, the fitted values satisfy \(\gamma>0\) and \(p<\beta\), illustrating the exponential suppression of the logical error rate with increasing code distance.
We further use these fits to estimate the required code distances to reach \(p_{\mathrm L}<10^{-9}\) at \(\Omega/2\pi=11.9\,\mathrm{MHz}\).
The one-sided \(95\%\) upper bounds fall below \(10^{-9}\) at \(d=13\), \(17\) and \(13\) for \(\bar H\), \(\bar S^2\) and \(\overline{\mathrm{CZ}}\), respectively.
Thus, a common code distance \(d=17\) is sufficient for all three logical Clifford gates to achieve \(p_{\mathrm L}<10^{-9}\) at \(\Omega/2\pi=11.9\,\mathrm{MHz}\).

\begin{samepage}
\section{Magic state cultivation}
\label{sec:si_magic_state_cultivation}

We describe the \(\mathsf{GRAIL}\) implementation of magic state cultivation and the decoding, postselection and statistical analysis used to evaluate its performance.

\end{samepage}

\subsection{Physical implementation of magic state cultivation}
\label{sec:msc_benchmark}

\begin{table}[htbp]
\centering
\small
\renewcommand{\arraystretch}{1.12}
\caption{\textbf{Stages of magic state cultivation in \(\mathsf{GRAIL}\).}
The last column lists differences from the \(f=5\) fold-transversal cultivation protocol of Ref.~\cite{sahay2026foldmsc}.}
\label{tab:msc_stage_sequence}
\begin{tabular}{@{}>{\raggedright\arraybackslash}p{0.15\linewidth}@{\hspace{0.02\linewidth}}>{\raggedright\arraybackslash}p{0.30\linewidth}@{\hspace{0.02\linewidth}}>{\raggedright\arraybackslash}p{0.51\linewidth}@{}}
\toprule
Stage & Description & Main differences \\
\midrule

1. Inject
& Unitary encoding of \(T\lvert+\rangle\) in \(\mathrm{Rot}(3)\), followed by one SE round.
& Rb-mediated Cs--Cs CZ gates are used during encoding.
After encoding and before SE, addressed \(H\) gates on some Cs data qubits implement the basis change from CSS to ZXXZ~\cite{xzzx2021,zxxz}. \\
\addlinespace

2. Grow
& Unitary growth from \(\mathrm{Rot}(3)\) to \(\mathrm{Reg}(3)\).
& After the \(\mathrm{Rot}(3)\) SE round and before growth, we apply a single layer of addressed \(H\) gates to two subsets of Cs atoms.
The gates on the first subset change the basis from ZXXZ to CSS, while those on the second subset prepare Cs qubits in \(\lvert+\rangle\) for code growth.\\ 
\addlinespace

3. Check \(\times 2\)
& Two fold-transversal \(\bar H_{XY}\) checks on \(\mathrm{Reg}(3)\).
& Parallelized GHZ preparation and CCZ implementation using a nine-atom Rb GHZ ancilla. \\
\addlinespace

4. Grow
& Unitary growth from \(\mathrm{Reg}(3)\) to \(\mathrm{Rot}(5)\).
& The Cs atoms to be incorporated into the enlarged \(\mathrm{Rot}(5)\) patch have been exposed to global pulses while idling in the array.
Before growth, we therefore reset Rb atoms to \(\lvert0\rangle\) and transfer their initialized states to these Cs atoms using SWAP gates.
We then apply addressed \(H\) gates to the Cs qubits that must start in \(\lvert+\rangle\). \\
\addlinespace

5. Stabilize \(\times 2\)
& Two rounds of stabilizer measurements on \(\mathrm{Rot}(5)\).
& Before the first SE round on \(\mathrm{Rot}(5)\), we apply addressed \(H\) gates to realize a basis change from CSS to ZXXZ. \\
\addlinespace

6. Grow
& Unitary growth from \(\mathrm{Rot}(5)\) to \(\mathrm{Reg}(5)\).
& Similar to stage~2.
\\

7. Check \(\times 2\)
& Two fold-transversal \(\bar H_{XY}\) checks on \(\mathrm{Reg}(5)\).
& Similar to stage~3.
\\

8. Grow
& Unitary growth from \(\mathrm{Reg}(5)\) to \(\mathrm{Rot}(9)\).
& Similar to stage~4.
\\
\addlinespace

9. Escape
& Direct growth from \(\mathrm{Rot}(9)\) to \(\mathrm{Rot}(13)\), followed by five SE rounds on \(\mathrm{Rot}(13)\).
& Additional Cs storage atoms, global Rb reset and a global Rb \(H\) are combined to prepare the \(\lvert0\rangle\) and \(\lvert+\rangle\) inputs required for growth.
We SWAP the additional data qubits from Rb to Cs and apply addressed \(H\) gates on selected qubits of the original \(\mathrm{Rot}(9)\) patch to complete the basis change from CSS to ZXXZ.
We protect the cultivated \(T\) state on \(\mathrm{Rot}(13)\) with SE rounds.
The basis change from ZXXZ to CSS is applied before delivering the encoded state. \\
\bottomrule
\end{tabular}
\end{table}

\begin{figure}[htbp]
\centering
\includegraphics[width=0.3\linewidth]{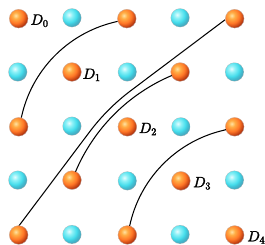}
\caption{\textbf{Gate layout for the logical \(\bar H_{XY}\) check.}
The \(\mathrm{Reg}(3)\) patch is shown, with Cs data atoms (orange) and Rb atoms (blue).
The five Cs data qubits \(D_0,\ldots,D_4\) lie along the fold diagonal:
\(U_+\) acts on \(D_0,D_2,D_4\), and \(U_-\) acts on \(D_1,D_3\).
The four black lines connect mirrored Cs pairs on which the CZ gates act.
Each of these nine gates in Eq.~\eqref{eq:msc_target_check_factors} is controlled by a distinct Rb atom of the $9$-atom GHZ ancilla, yielding $3$ controlled-\(U_+\) gates, $2$ controlled-\(U_-\) gates and $4$ CCZ gates.
For \(\mathrm{Reg}(d)\), the corresponding number of GHZ ancilla atoms is \(d+(d-1)+(d-1)^2=d^2\).}
\label{fig:msc_logical_check_factors}
\end{figure}

Building on the sequence described in the main text (Fig.~5\textbf{a}) and Methods, we detail the controlled-\(\bar H_{XY}\) operation following the two CZ layers of the LOCC-based construction, together with the subsequent postponed measurements and feedback, for cultivation on \(\mathrm{Reg}(d)\), with \(d=3,5\).
On each unrotated surface code \(\mathrm{Reg}(d)\) patch, with \(d=3,5\), we label the Cs data qubits along the fold diagonal by \(D_0,D_1,\ldots,D_{2d-2}\) and denote the set of Cs pairs related by reflection across this diagonal by \(\mathcal Q_{\mathrm{mir}}\).
The physical operator implementing \(\bar H_{XY}\) on the code space factorizes as~\cite{sahay2026foldmsc}
\begin{equation}
U_{\bar H_{XY}}
=
\prod_{i\,\mathrm{even}} (U_+)_{D_i}
\prod_{i\,\mathrm{odd}} (U_-)_{D_i}
\prod_{(a,b)\in\mathcal Q_{\mathrm{mir}}}\mathrm{CZ}_{a,b},
\qquad
U_+=\frac{X+Y}{\sqrt{2}},
\quad
U_-=\frac{Y-X}{\sqrt{2}}.
\label{eq:msc_target_check_factors}
\end{equation}
This decomposition contains \(d\) single-qubit \(U_+\) gates, \(d-1\) single-qubit \(U_-\) gates, and \((d-1)^2\) CZ gates, giving \(d^2\) gates in total. We assign one Rb GHZ ancilla atom to control each gate: the first \(2d-1\) atoms control the \(U_\pm\) gates along the fold diagonal, while the remaining \((d-1)^2\) atoms control the CZ gates on mirrored pairs, forming CCZ gates.
Supplementary Fig.~\ref{fig:msc_logical_check_factors} illustrates the diagonal single-qubit gates and mirrored-pair CZ gates for \(\mathrm{Reg}(3)\).

Using the gate constructions in Supplementary Note~\ref{sec:physical_gate_decompositions}, we apply the controlled-\(U_+\) gates in parallel, followed by the controlled-\(U_-\) gates in parallel and then the CCZ gates in parallel.
The single-qubit gate parameters are
\begin{equation}
(\mu_+,\alpha,\varphi)
=\left(\frac{\hat{x}-\hat{y}}{\sqrt{2}},\frac{\pi}{2},\pi\right),
\qquad
(\mu_-,\alpha,\varphi)
=\left(\frac{\hat{x}+\hat{y}}{\sqrt{2}},\frac{\pi}{2},\pi\right).
\end{equation}
After the controlled operations, we apply \(G_{\mathrm{Rb}}(H)\) followed by global Rb readout \(\mathcal M_{\mathrm{Rb}}\), measuring both the GHZ ancillas and the additional Rb ancillas used for GHZ preparation in the \(X\) basis. 
Let \(m_{\mathrm{cat},j}\) and \(m_{\mathrm{anc},j}\) denote their respective measurement bits. We define the logical measurement outcome and GHZ-preparation parity as
\begin{equation}
m_{\bar H_{XY}}=\bigoplus_{j=0}^{d^2-1}m_{\mathrm{cat},j},
\qquad
m_{\mathrm{GHZ}}=\bigoplus_{j=0}^{d^2-1}m_{\mathrm{anc},j}.
\label{eq:msc_check_parities}
\end{equation}
Here, \(m_{\bar H_{XY}}\) gives the logical \(\bar H_{XY}\) measurement outcome and \(m_{\mathrm{GHZ}}\) checks the GHZ preparation. 
The logical check passes postselection only when both parities vanish.

We next describe how feedback from the LOCC-based GHZ preparation is postponed in our implementation~\cite{Piroli2021LOCC}.
For the $k$-th Rb atom of the GHZ state and the unitary $U_k$ it controls in Eq.~\eqref{eq:msc_target_check_factors}, the postponed correction \(X_k\) from LOCC propagates through the controlled-$\bar H_{XY}$ gate \(\mathrm {CU}_{\bar H_{XY}}\) as
\begin{equation}
\mathrm {CU}_{\bar H_{XY}}\,(X_k \otimes I)
=(X_k \otimes U_k)\,\mathrm {CU}_{\bar H_{XY}}.
\end{equation}
A complete logical \(\bar H_{XY}\) check stage on \(\mathrm{Reg}(d)\) consists of two consecutive measurement rounds. 
We retain an attempt only if \(m_{\bar H_{XY}}=m_{\mathrm{GHZ}}=0\) in both rounds. Let \(m_{\mathrm{anc},j}^{(r)}\) denote the preparation-ancilla measurement outcomes in round \(r=1,2\). The corresponding feedback bits are defined as
\begin{equation}
\mathrm{fb}_0^{(r)}=0,\qquad
\mathrm{fb}_k^{(r)}
=\bigoplus_{j=0}^{k-1}m_{\mathrm{anc},j}^{(r)},
\quad 1\leq k<d^2.
\end{equation}
Rather than applying feedback after each round, we combine the correction bits as
\begin{equation}
\mathrm{fb}_{\mathrm{tot},k}
=\mathrm{fb}_k^{(1)}\oplus\mathrm{fb}_k^{(2)}.
\end{equation}
If both rounds pass postselection, we apply the combined feedback by performing \(U_k\) on the corresponding Cs qubit or mirrored pair whenever \(\mathrm{fb}_{\mathrm{tot},k}=1\), before proceeding to the next growth stage.

\subsection{Decoding, postselection and statistical estimation}
\label{sec:msc_statistics}

\begin{figure}[htbp]
\centering
\includegraphics[width=\linewidth]{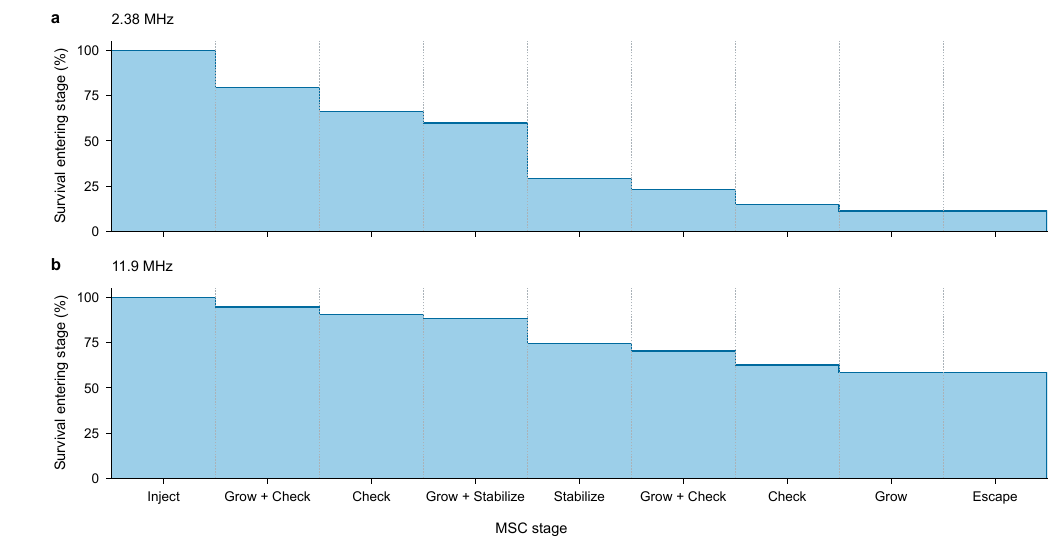}
\caption{\textbf{Stage survival in the \(f=5\) cultivation benchmark.}
\textbf{a} and \textbf{b} show the results at \(\Omega/2\pi=2.38\,\mathrm{MHz}\) and \(11.9\,\mathrm{MHz}\), respectively.
Each bar gives the percentage of initial attempts that survive all preceding checks and enter the indicated stage.
At the two frequencies, \(11.2\%\) and \(58.4\%\) of attempts pass all hard postselection checks, respectively.
}
\label{fig:msc_stage_survival_active_extremes}
\end{figure}

We use the logical-\(\bar Y\) cultivation benchmark to estimate the logical error rate and the expected number of attempts per accepted output in Fig.~5\textbf{b} of the main text~\cite{sahay2026foldmsc}.
An attempt passes hard postselection if it satisfies all stabilizer and logical-check parity conditions.
Supplementary Fig.~\ref{fig:msc_stage_survival_active_extremes} shows the accumulated survival rate at each stage under the hardware-inferred circuit-level noise model.

For attempts that pass hard postselection, we use the complementary gap obtained from the \(\mathrm{Rot}(13)\) syndrome record to assess the reliability of the decoded logical outcome and retain only attempts with a gap at or above the chosen cutoff~\cite{2024msc,sahay2026foldmsc,Gidney2025Yoked,Bombin2024FTPostselection,Hutter2014SurfaceCode,Meister2024SoftOutput}.
At each cutoff, let \(N_{\mathrm{attempt}}\) be the total number of attempts, \(N_{\mathrm{kept}}\) the number of attempts passing both hard postselection and the gap cutoff, and \(N_{\mathrm{error}}\) the number of logical errors among the retained attempts.
The estimated logical error rate and the expected number of attempts per accepted output are
\begin{equation}
\widehat{\varepsilon}_L=\frac{N_{\mathrm{error}}}{N_{\mathrm{kept}}},
\qquad
\frac{1}{p_{\mathrm{final}}}=\frac{N_{\mathrm{attempt}}}{N_{\mathrm{kept}}},
\label{eq:locc_gap_estimators}
\end{equation}
where \(p_{\mathrm{final}}\) is the fraction of attempts retained after both postselection steps.

The shaded bands in Fig.~5\textbf{b} of the main text are two-sided \(95\%\) confidence intervals calculated from \(N_{\mathrm{error}}\) logical errors in \(N_{\mathrm{kept}}\) retained attempts at each fixed cutoff.
For each Rabi frequency, we select the cutoff that minimizes \(\widehat{\varepsilon}_L\).
Supplementary Table~\ref{tab:msc_operating_points} gives the counts and intervals at the selected cutoffs for the two Rabi frequencies highlighted in the main text.

\begin{table}[htbp]
\centering
\caption{\textbf{Logical error rates at the two reported operating points.}
The \(95\%\) confidence interval for each logical error rate treats the selected gap cutoff as fixed.}
\label{tab:msc_operating_points}
\small
\setlength{\tabcolsep}{4pt}
\begin{tabular}{@{}
>{\centering\arraybackslash}m{0.10\linewidth}
>{\centering\arraybackslash}m{0.11\linewidth}
>{\centering\arraybackslash}m{0.22\linewidth}
>{\centering\arraybackslash}m{0.14\linewidth}
>{\centering\arraybackslash}m{0.23\linewidth}
>{\centering\arraybackslash}m{0.10\linewidth}
@{}}
\toprule
\shortstack{\(\Omega/2\pi\)\\(MHz)} &
\shortstack{Gap cutoff\\(dB)} &
\(N_{\mathrm{error}}/N_{\mathrm{kept}}\) &
\(\widehat{\varepsilon}_L\) &
\shortstack{\(95\%\) confidence\\interval} &\shortstack{Expected\\attempts} \\
\midrule
\(2.38\) & \(110\) & \(67/345\,435\,707\) & \(1.94\times10^{-7}\) & \([1.50,\,2.46]\times10^{-7}\) & \(258\) \\
\(11.9\) & \(118\) & \(5/25\,246\,683\,992\) & \(1.98\times10^{-10}\) & \([0.643,\,4.62]\times10^{-10}\) & \(2.43\) \\
\bottomrule
\end{tabular}
\end{table}

\FloatBarrier
\renewcommand{\refname}{Supplementary References}
\phantomsection
\addcontentsline{toc}{section}{Supplementary References}
\bibliographystyle{sn-nature}
\bibliography{references_ncs}